\documentclass[useAMS,usenatbib]{mn2e}
\usepackage{graphicx}
\begin{document}
\title[Foreground contributions to the Cosmic Microwave Background]{Foreground contributions to the Cosmic Microwave Background}

\author[T. Wibig and A. W. Wolfendale]{T. Wibig $^{1}$\thanks{E-mail:
wibig@zpk.u.lodz.pl;} and A. W. Wolfendale$^{2}
$\\
$^{1}$Experimental Physics Dept., University of \L \'{o}d\'{z};
The Andrzej So\l tan Institute For Nuclear Studies,\\
Cosmic Ray Lab., \L \'{o}d\'{z}, Uniwersytecka 5,
POB 447, \L \'{o}d\'{z} 1; Poland\\
$^{2}$Physics Department, University of Durham, Durham DHI 3LE, UK.}



\maketitle


\begin{abstract}

A detailed search has been made for evidence of residual
foreground contributions to the Cosmic Microwave Background (CMB)
WMAP \citep{1,30}, 
 a map (nominally)
cleaned for foreground already \citep{3}.
We find positive results in that
various features relate to Galactic properties. For example, on
the largest angular scales we find significant differences between
the power in the fluctuations for positive and negative Galactic
latitudes and between the four Galactic Quadrants.  There are also
differences between the power spectrum at latitudes within
10$^\circ$ of the Plane and at higher latitudes.  The mean
temperature shows similar variations. An explanation in terms of
Galactic effects seems inescapable.

In an effort to find the origin of these Galactic-style effects we
have examined the evidence from Galactic gamma rays, specifically
from the EGRET instrument \citep{4}.  We are mindful that the CMB
maps examined \citep{3}
have already been `cleaned' (for
cosmic ray and other effects) in a rather complex way but, in our
view, the cleaning has left potentially serious 'contaminations'.
A correlation is found between gamma ray intensities and the CMB
and other cosmic ray indicators. For example, regions of the
Galaxy having (line of sight) steep cosmic ray energy spectra have
low mean CMB temperatures and the important Loop I edge region,
where the cosmic ray intensity is high, has a high mean
temperature.

Most of the large scale Galactic asymmetries (eg North, South
difference and Quadrant variations) have analogues in cosmic ray
asymmetries and also in some other Galactic properties, such as
the column density of gas.  Thus, it is possible to hypothesize on
direct cosmic ray-induced contributions, although it may be that
cosmic rays are simply the indicators of Galactic `conditions'
which are influencing the residual CMB fluctuations.

Irrespective of the actual cause of the correlations we have
endeavoured to extrapolate to the situation where the residual
foreground is minimised.  The effect on the usually derived
cosmological properties is briefly examined.  The least that can
be said is that the 'error' in some of these properties has been
underestimated.
\end{abstract}


\section{Introduction}
\label{i}
It is well known that although the Galactic foreground to the CMB
measurements should be small, it is not zero.  Some 15 years ago
we \citep{5,6}
 drew attention to the fact that, despite the contribution to
the `CMB' spectrum from electron bremsstrahlung falling rapidly
with frequency, the contribution from dust rises rapidly and there
is a frequency region -- from 30 - 100 GHz - where their sum has
nearly the same 2.7K spectrum. The derived value of $\Delta T/T$
was about $2.5\times 10^{-6}$, ie $\Delta T$ is about 7$\mu$K. The
angular scale was 8$^\circ$ relating to the Tenerife experiment
\citep{7}
(beam width 8.8$^\circ$ FWHM and the beam throw 8.2$^\circ$), and
the value related to a `quiet' region of the Galaxy, 'quiet' in
the sense of having low radio and dust signals. The corresponding
value for $b= 90^\circ$ was close to this value. It was claimed,
further, that the minimum foreground (synchrotron plus dust) would
be at a frequency of 70 GHz for the 'DC signal', ie the absolute
intensity as distinct from the fluctuations in the foreground
temperature.
 An important parameter is the degree of
flatness of the foreground temperature vs frequency.  Taking a
straight line between 20 and 90 GHz as datum, the maximum downward
deviation was a factor 3.0. Actual measurements \citep{14} now
enable the DC level to be determined. WMAP data at the five
frequencies (22.8 to 93.5 GHz) do indeed show a minimum signal at
about 60 GHz and the form of the dependence of temperature on
frequency is even flatter, the factor being only 2.0.  The
absolute temperature level is somewhat higher, even, than 'our'
early prediction.

Taking our predicted temperature fluctuations, as distinct from
the DC level,
at $\Delta T \sim 7 \mu K$ for the
minimum, the average over much of the Galaxy, but away from the
Galactic Plane, would be of order 20$\mu K$.  Inspection of the
power spectrum for the WMAP \citep{30} shows that for 8$^\circ$
($\ell\sim 36$) the average power is $\sim 1000$ $\mu K^{2}$.
'Our' 400 $\mu K^{2}$ is therefore a significant fraction.  There
is thus a prima facie case for a serious contribution.

Correction of the WMAP for the Galactic foreground has been made
by the WMAP group \citep{8} who have carried out a procedure to
generate, not a published map but a power spectrum. Many workers
have used this power spectrum to derive important cosmological
parameters. Here we will use a cleaned map \citep{3}
the cleaning of which
was carried out by a more sophisticated technique, specifically
the individual frequency sets of data were weighted not just
according to frequency (as in \cite{8}) but also by harmonic
($\ell$) in the power spectrum analysis.

An important point to be mentioned at this stage is the fact that
the well-known cosmic ray (electron synchrotron) contribution to
the foreground at low frequencies is correlated with that of dust
at the highest frequencies.  This is a consequence of the
well-known strong correlation between the synchrotron and far
infra-red radiation emission of galaxies, commented on by many
workers over the years, including ourselves eg \citep{10}. Although
there can be no certainty about the mechanism, a possibility
relates to CR-heating of the very cold dust clouds in the Galactic
Halo.  In this respect we are mindful of the comment \citep{11}
that "... between a few $K$ and about 30 $K$, the (neutral) gas is
almost entirely in the form of molecular hydrogen, residing in
dense clumps...".  Interestingly, confining attention to cold
clumps, their contribution could well fall off with Galactic
latitude more slowly than cosec~$|b|$, because many of the low
$|b|$ clouds would be warmed by Galactic Plane radiation and not
recorded in the CMB map. In fact, the $b$-dependence may not be at
all simple.

Before continuing it is instructive to examine the various energy
densities.  The CMB radiation has an energy density of 0.24
eV$\:$cm$^{-3}$ and is universal; that in cosmic rays is $\sim
0.5$ eV$\:$cm$^{-3}$, and, as is well known, this is the same as
that in interstellar magnetic fields, gas motion and starlight --
all in the local region of the Galaxy. The fluctuations in the CMB
correspond to about (1-10)$\times 10^{-6}$ eV$\:$cm$^{-3}$ -- on
various angular scales -- and a fraction of the CR energy density
(or of one of the other components of similar energy density) of
order 10$^{-6}$ going into pseudo-CMB signals would not appear
impossible.  In view of the importance of the CR `background' in
other searches, eg for cosmic neutrinos, dark matter particles,
magnetic monopoles, 
 it would perhaps be surprising
if CR effects were absent.

How much of the CR foreground has in fact been
taken out by the `cleaning' is open to question but we presume
that most of the contribution has been removed and that the
residual will vary slowly with Galactic latitude. Insofar as the
synchrotron radiation has a pronounced asymmetry about the
Galactic Plane, with a North/South (N/S) ratio less than unity
\citep{8,11} -- and all the WMAP frequencies show this effect
before correction \citep{8} we will search for such an asymmetry
in the cleaned map; such an asymmetry is also present in other
Galactic phenomena, as will be demonstrated later. Similarly, in
view of all Galactic phenomena being dependent on the Galactic
Quadrant concerned, Quadrant variations will also be searched for.

Our work is not the first to claim asymmetries in the cleaned
WMAP.  Although the WMAP group 
\citep{30,8}
 claimed that there is
no evidence for non-Gaussian primordial fluctuations in the WMAP,
and they quote many other workers who have come to the same
conclusion, other recent studies \citep{12},
 have shown that 'strong
non-Gaussian features emerge when considering the Northern and
Southern Galactic hemispheres separately'.  Others still
\citep{32,15}
 had, previously,
found asymmetries in the distribution of the CMB fluctuations in
the two hemispheres. Thus, the present work is another
contribution to the debate on asymmetries in general, but with an
independent approach.

The cosmic ray aspect is, so far as we know, truly original.

Some remarks are necessary first about the adopted map used
throughout the analysis.  This is the 'high resolution foreground-
cleaned CMB map from WMAP' given by 
 \cite{3}. Although the
WMAP group itself has analysed the basic data from the WMAP
instrument in great detail, and derived best-fit results for the
various cosmological parameters \citep{1,30},
it has not given the
necessary cleaned-map data. Furthermore, its treatment of the five
WMAP frequency maps to allow for foreground appears to us to be
rudimentary in that $\ell$-independent statistical weights were
applied before the data were combined. In the map adopted by us,
\citep{3}, $\ell$-dependent
weights were used and, although we contend that significant
foreground remains, the data should be better.

We estimate that the resolution of the adopted data is about 12'
and that, although the minimum in the power spectrum at $\ell \sim 350$
is enhanced somewhat, the effect on the 'one-degree-peak'
should be small.  The authors state that the Galactic residuals
seen along the Galactic Plane 'are present in only a tiny fraction
of the total sky area and therefore contribute little to the total
power spectrum'.

\section{Large scale asymmetries} \label{ii}
\subsection{The Parameters Under Study }\label{iiA}

The CMB sky is characterized by the temperatures of individual
pixels and these can be studied in a variety of ways, as follows:
\begin{itemize}
\item[i)] The search for discrete excesses and deficits.  For
sizes of the order of a few pixels most will relate to
extragalactic systems such as active galactic nuclei.  We have
identified several AGN and the VIRGO, PEGASUS 1 and PISCES
clusters.
  In general, we find six times as many
excesses as we find from searches in 'artificial universes', taken
from the conventional CMB power spectrum.  Turning to deficits (eg
Sunyaev-Zeldovich reductions in galaxy clusters) we find twice as
many in the genuine map as in artificial universes.  The
significance of the above is two fold; our technique of analysis
receives a measure of support; the integrated magnitude of their
contributions is small - and only affects the results at the
highest of $\ell$ values.

\item[ii)] An analysis of the power spectrum formed from the basic
data.  The standard technique is usually adopted to give
$W(\ell)=\ell\: (\ell+1)\:C_\ell/ 2 \pi$,
 where $\ell$ is the
multipole
and $C_\ell$ is the angular power spectrum average derived by
averaging the sum of the squares of the terms in the spherical
harmonic. The expression gives the power per
  logarithmic intervals of $\ell$ and inspection
of the form of $W(\ell)$, eg \citep{30,8,3},
 shows that much of
the power is at large $\ell$, specifically in the region of the
'one-degree peak'. We expect the Galactic foreground effects to be
on the scales of rather small $\ell$ (perhaps $\ell=10$) as
distinct from the $\ell=200$ related to the 'one-degree peak' but
all scales are, in principle, possible. Here, we usually study the
$\ell$-dependence of parameters but, where undue complexity would
arise, we give results for just $\ell=10$ and $\ell=100$.

\item [iii)] Averaging the sky over large regions gives the mean
CMB temperature,$\langle\Delta T \rangle_{\rm CMB}$.  By definition, over the whole
sky this will be zero, but for restricted regions the value will
be finite, and its study can be useful. Inspection of the
$\Delta T_{\rm CMB}$ distribution for individual pixels shows a median
excess of 80 $\mu K$ (of those which are possible) with 10\%
 having $\Delta T > 200\: \mu K$.  The 80 $\mu K$ level in the
standard WMAP maps is the 'green level', with many regions having
angular size of order several tens
of degrees.
\end{itemize}
From what has been stated it is apparent that it will be valuable
to look at features in the map from both the power -- and the mean
temperature -- points of view.

\subsection {North-South Asymmetry} \label{iiB}
For all Galactic radiations there is a major symmetry, with
respect to latitude, together with a
 N-S asymmetry (details of which are given
later). The former usually predominates but the effect of the map
cleaning would be expected to remove most of this; the result is
that the symmetry and asymmetry may be of not too dissimilar
magnitude.  A near cosec $|b|$ dependence (but bearing in mind
previous remarks) might be expected; in fact, any disk-like
emission (or absorption) would be expected to give such a
$|b|$-dependence, assuming near constant emissivity per unit
volume over the Galactic disc.
  It should be remarked that the
original WMAP results for the individual frequencies \citep{8},
show a pronounced cosec $|b|$ dependence over the range for which
data are given: cosec$|b|$ = 1 to 4, the slope of which is
systematically higher in the South than the North (there are also
subtleties in the results to which we will refer later).  Thus,
there are both symmetries and asymmetries.  A basic question is
"to what extent has the cosec $|b|$ dependence, and with it the
largest scale Galactic foreground, been taken out in the
'cleaning-process'?"

\begin{figure}
 \centerline{
\includegraphics[width=9cm]{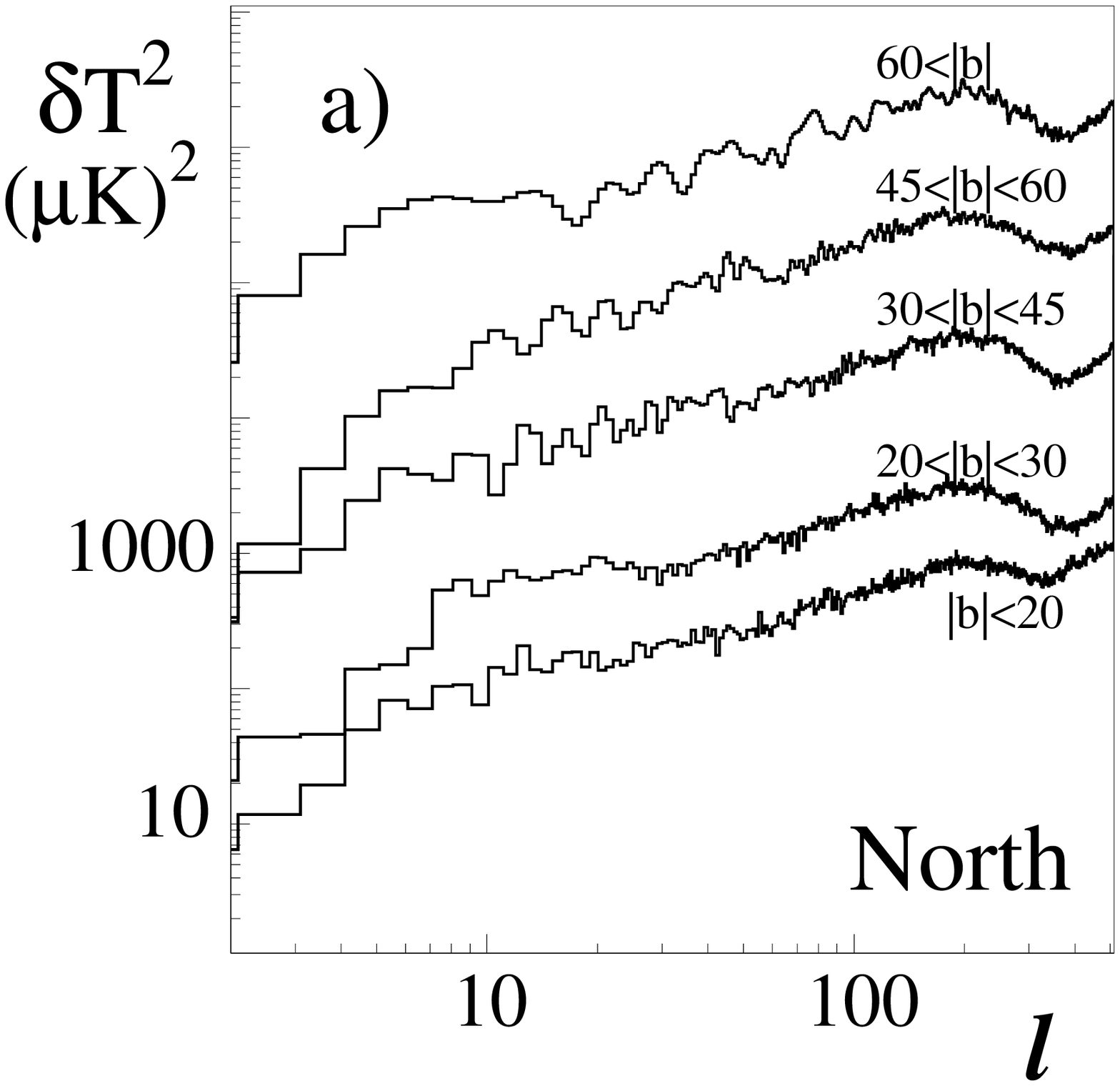}
\includegraphics[width=9cm]{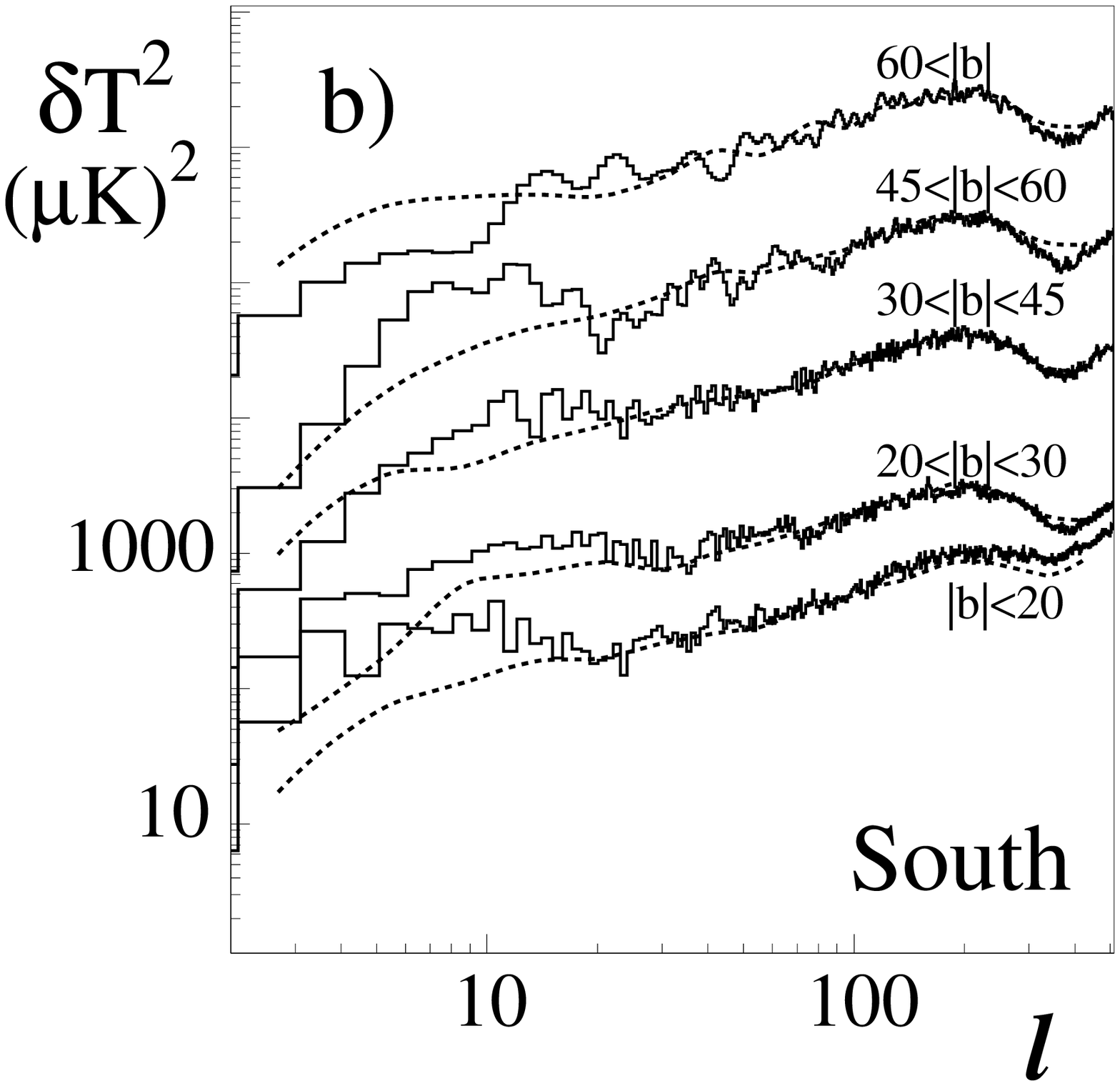}  }
 \centerline{
\includegraphics[width=9cm]{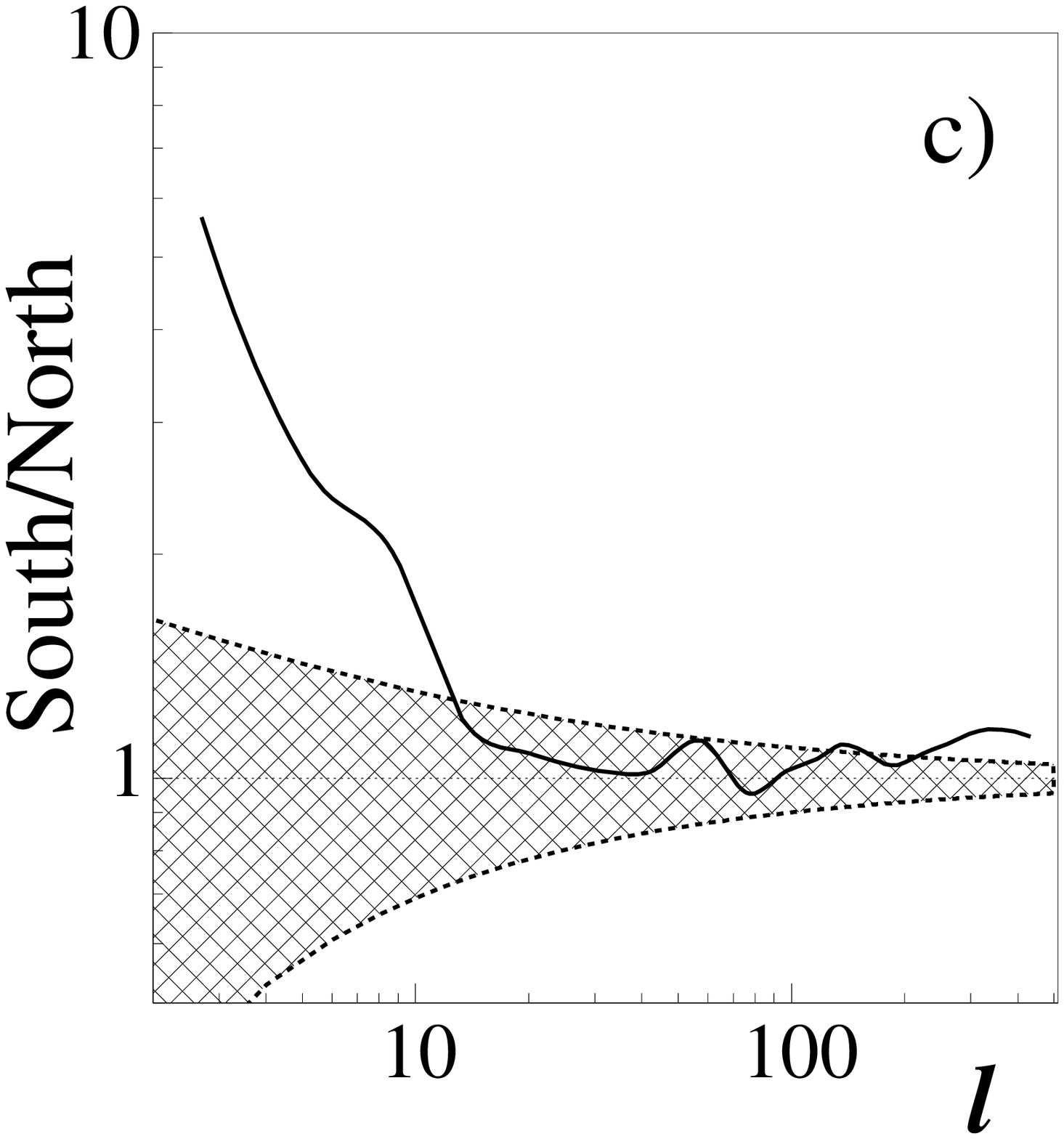}
\includegraphics[width=9cm]{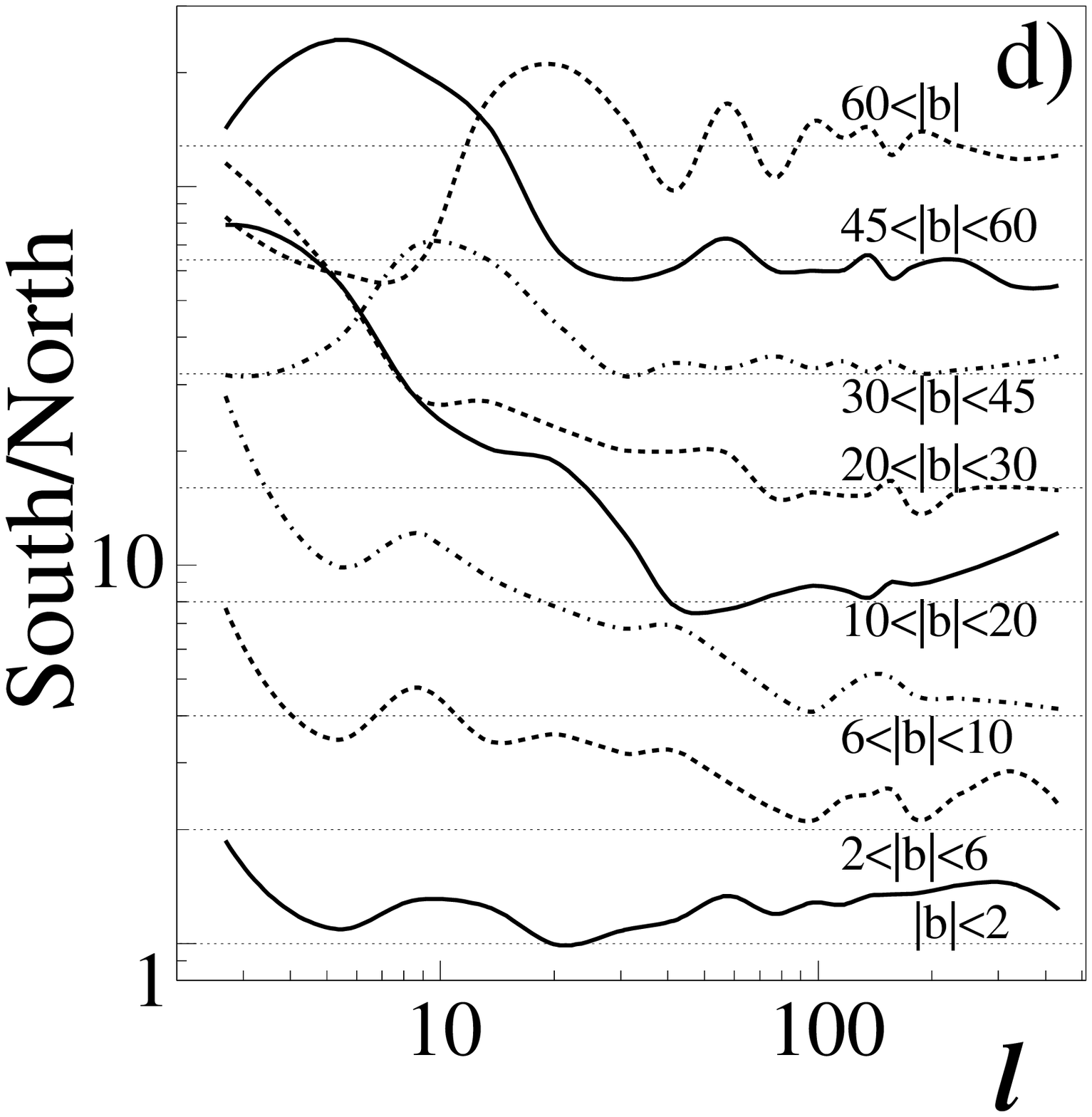} }
\caption{Power spectrum of the WMAP data
\citep{3}
 for the two Galactic hemispheres
(a) North and
(b) South.
 Moving upwards, 
  the latitude ranges are (plus or
minus) :
$<20^\circ$, $20^\circ/30^\circ$, $30^\circ/45^\circ$,
$45^\circ/60^\circ$ and $>60^\circ$.
The data for positive latitudes
(the North) have been smoothed and also plotted in (b) to allow
observation of the fact that the `Southern Excess' continues to
`high' latitudes. Successive spectra have been displaced upwards
by a factor 10 to aid appreciation.
(c) Power spectrum South/North ratio vs $\ell$ for all the data;
the 'cosmic variance' is also shown.
(d) South/North ratio for the
various latitude bands (there is a smoothing due to
interdependence at low latitudes).
 \label{f1}}

\end{figure}

The fact that the cosec $|b|$ dependence, and N-S asymmetry in the
basic WMAP data, are independent of frequency is an important
result in that it suggests the fact mentioned in Sect.~\ref{i}
that there may be a spatial connection between the 'low' frequency
emission (22 GHz), which is related to cosmic ray electrons, and
the 'high' frequency emission (93 GHz), which is affected by
Galactic dust. This was the logic behind adding the foreground
$\Delta T$ values from these two mechanisms in our earlier work.
\citep{5,6}. 
 We start by comparing the power spectra for the N \& S for the same
latitude bands.  We are mindful of the 'cosmic variance' which
pervades the interpretation of all the WMAP data, but particularly
at low values of $\ell$.
Figure ~\ref{f1} shows the situation.  The smooth curves through
the points for the North have been given in the graph for the
South, for illustration purposed only.  Interpretation needs to
allow for the non-independence of nearby $\ell$-values for a
particular set of data; we estimate that this dependence exists
over a range in $\ell$ of $\Delta \ell$ a factor 2.

Furthermore, there is the diminishing effect on the South/North
ratio at small $|b|$ and low $\ell$; for example, at $\ell=10$
($\sim 18^\circ$) the ratio should fall at $|b|<10^\circ$ where
the innermost $b$-values are separated by less than 20$^\circ$.  It does; from
3.1 at $|b| = 10^\circ - 20^\circ$ to 1.3 for $|b|<2^\circ$.

The most important plot is that of the South/North ratio vs
$\ell$.  Figure 1c shows the result summed over all latitudes;
also shown is the mean spread for cosmic variance alone. It is
interesting to note that the S-excess persists over virtually the
whole range of $\ell$.  Formally, it is most significant at the
largest $\ell$-values, a surprising result.  The statistical
significance of the excess is not easy to estimate but, allowing
for the dependence of nearby values (within
$\Delta \ell \sim$ a factor of 2), it is at the 1\% level.

Turning to the $b$-dependence of the S-N ratio, the significance
is far less than might be expected from the large excesses seen in
Figure~\ref{f1}d, because of the small solid angles for the
individual $b$-ranges.  Furthermore, the excesses are greatest at
low $\ell$ where lack of independence of adjacent bins,  for small
$b$, are most serious, as remarked already. However, the
consequent errors are not large; inspection of Figures \ref{f1}c
and \ref{f1}d
shows that the weighted (by solid angle) ratios from
Figure~\ref{f1}a
 are 3.14 ($\ell$=5) and 2.1($\ell$=10), to be
compared with 2.7 and 1.7 respectively.  The 'error' is thus only
about 20\%.

Restricting attention to $\ell<10$ and $|b|>10^\circ$, the
dependence is less serious and something useful can be said
Allowing for the reduced solid angles, the excesses are in the
region of the 'upper limit' to the variance for
$10^\circ<|b|<60^\circ$, a region having 70\% of the solid angle.
The corresponding significance of the southern excess, after
allowing for the fact that the measured ratio rises at a faster
rate than the upper limit expected by chance, gives a
formal chance probability of
 $\sim$3\%, this value relating to the
low $\ell$-values (much of the weight for the previous 1\% came
from higher $\ell$-values).

The results on the N/S ratio suggest that there is a finite
Galactic foreground, affecting both low $\ell$
(all $b$)
- and, indeed, perhaps all $\ell$.  In the
next section we continue to study the evidence for foreground
effects by examining the data Quadrant by Quadrant (Quadrant 1
covers the range $\ell=0^\circ$ to 90$^\circ$; Quadrant 2, the
range 90$^\circ-180^\circ$, and so on.  Quadrants 1 and 4 comprise
the Inner Galaxy and Quadrants 2 and 3 the Outer Galaxy).

\subsection{Quadrant-dependent effects} \label{iiC}

There are well-known differences in all Galactic properties,
Quadrant to Quadrant. Near the Galactic Plane, at least, the Inner
Galaxy has higher intensities of every radiation than the Outer
Galaxy. We start with the power spectrum.

\begin{figure}
\centerline{
\includegraphics[width=8cm]{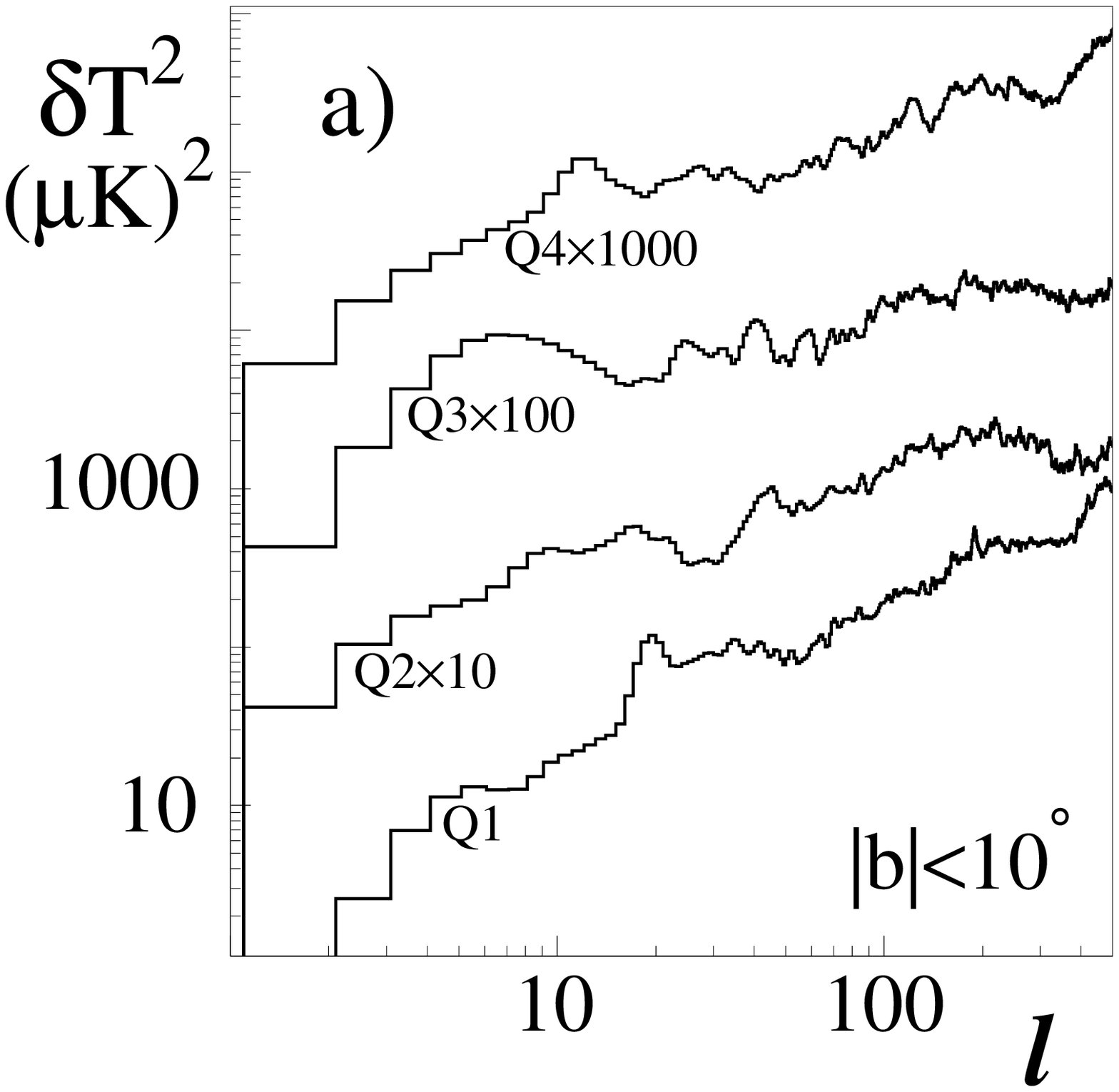}
\includegraphics[width=8cm]{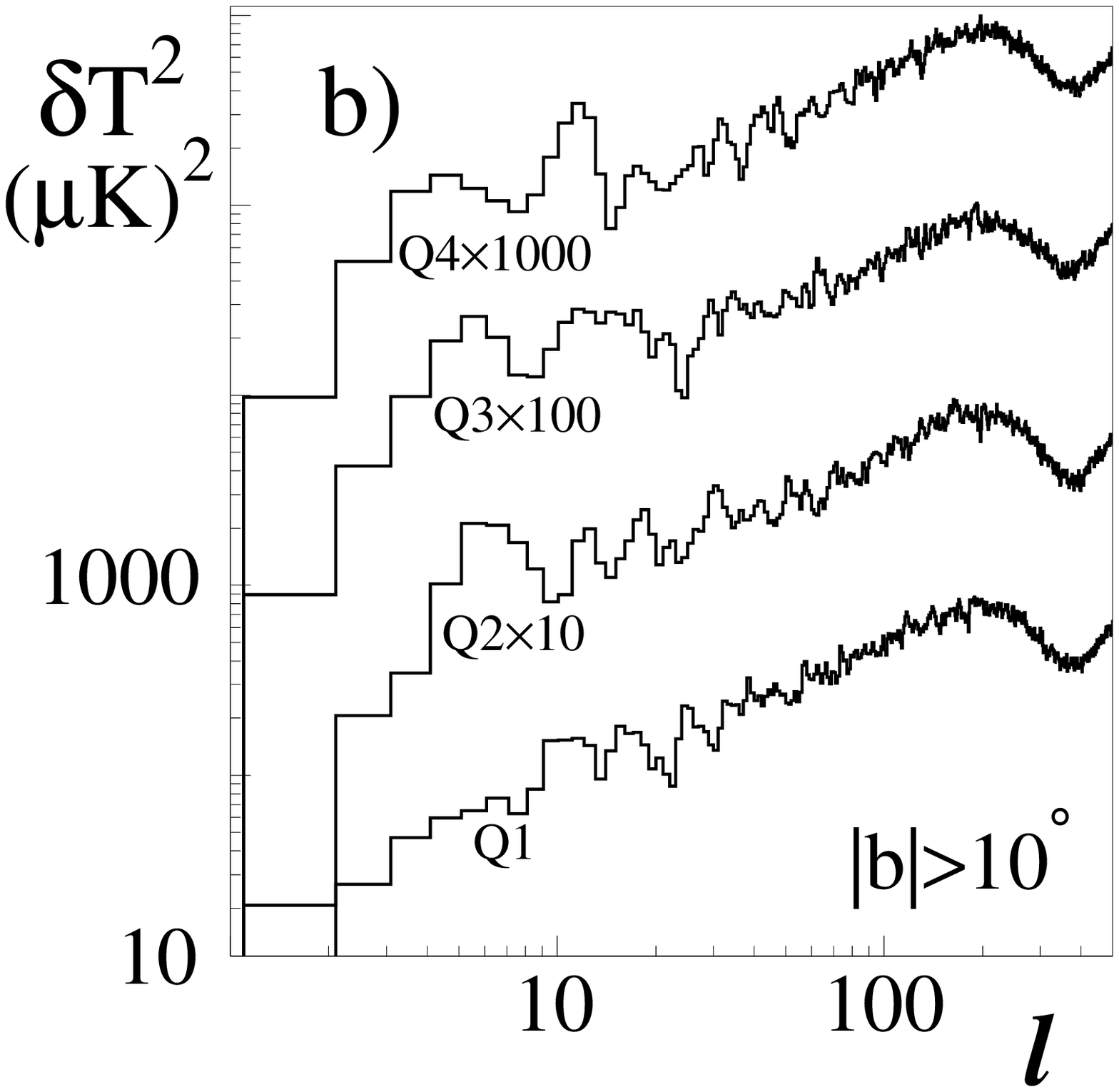}}

\vspace{-1cm}
\centerline{
\includegraphics[width=8cm]{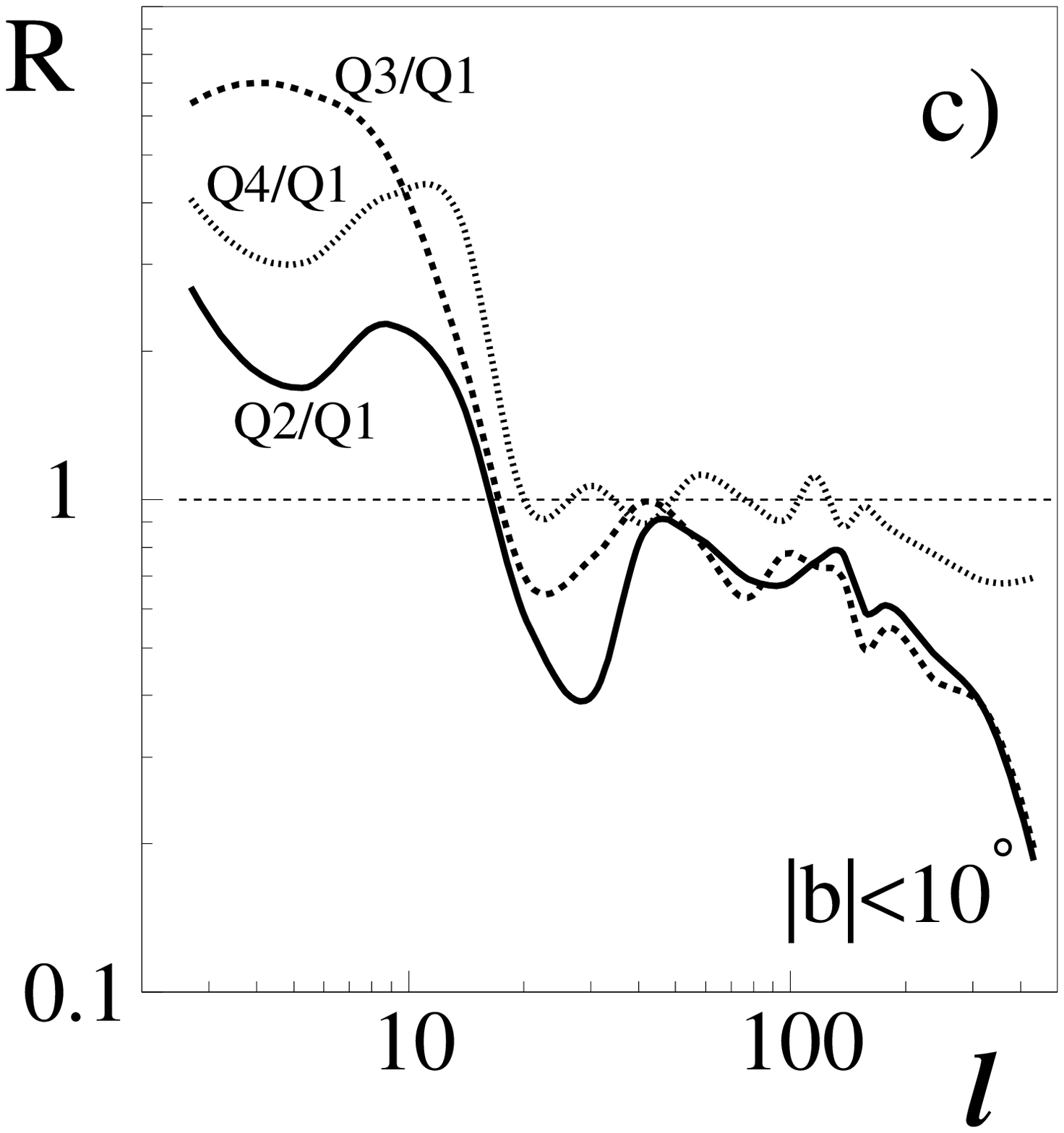}
\includegraphics[width=8cm]{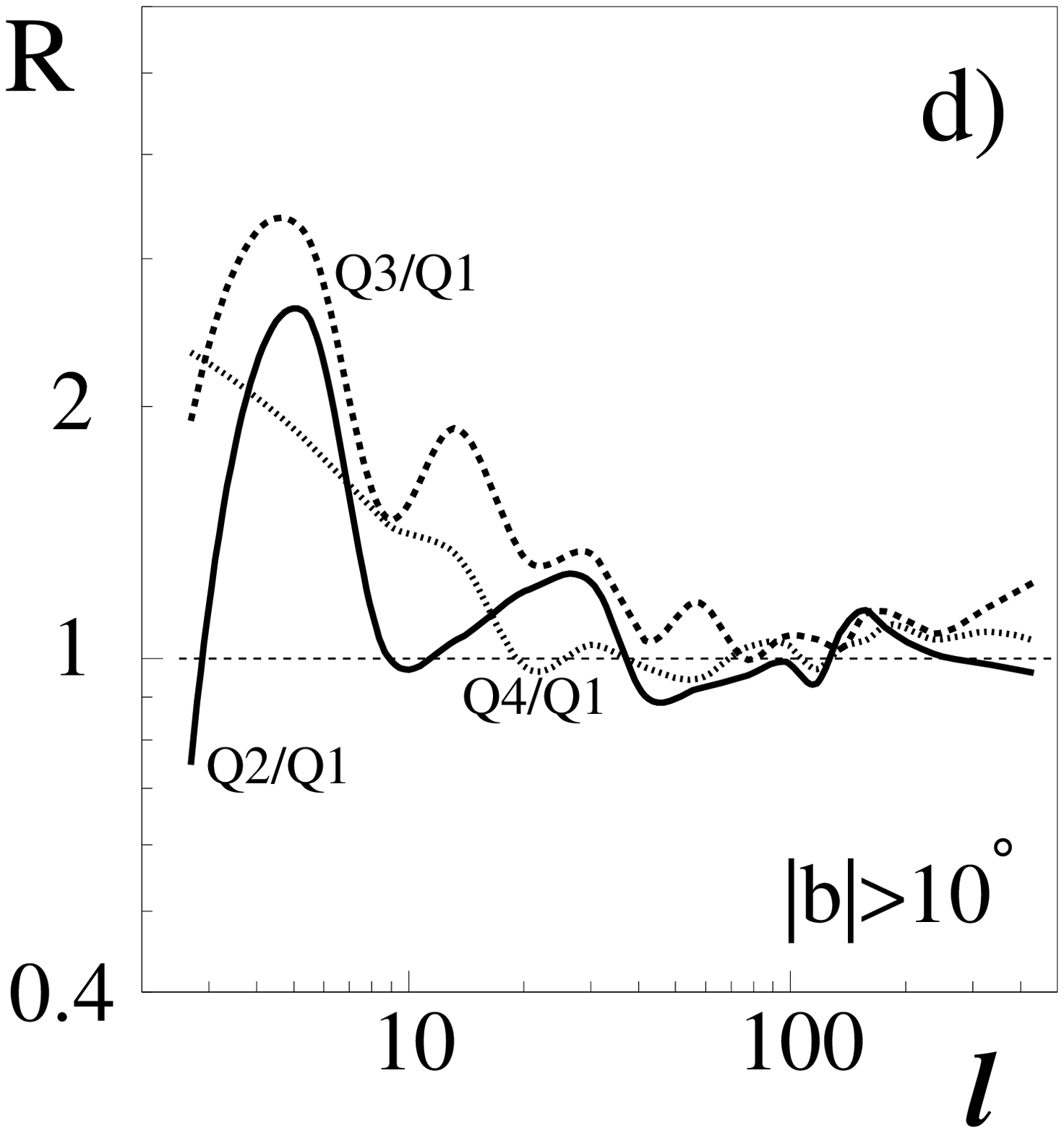}}

\vspace{-1cm}
\centerline{
\includegraphics[width=8cm]{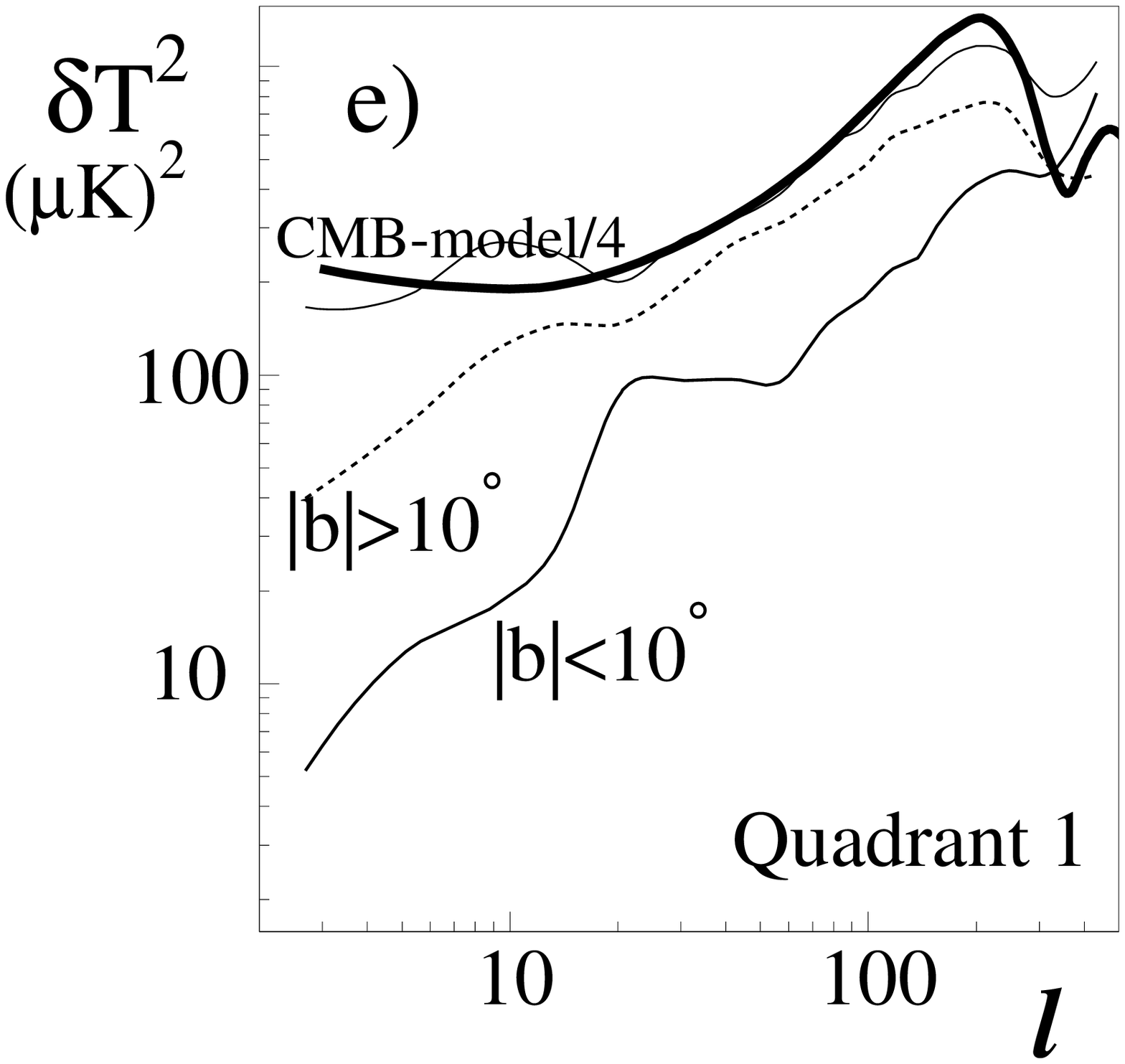}
\includegraphics[width=8cm]{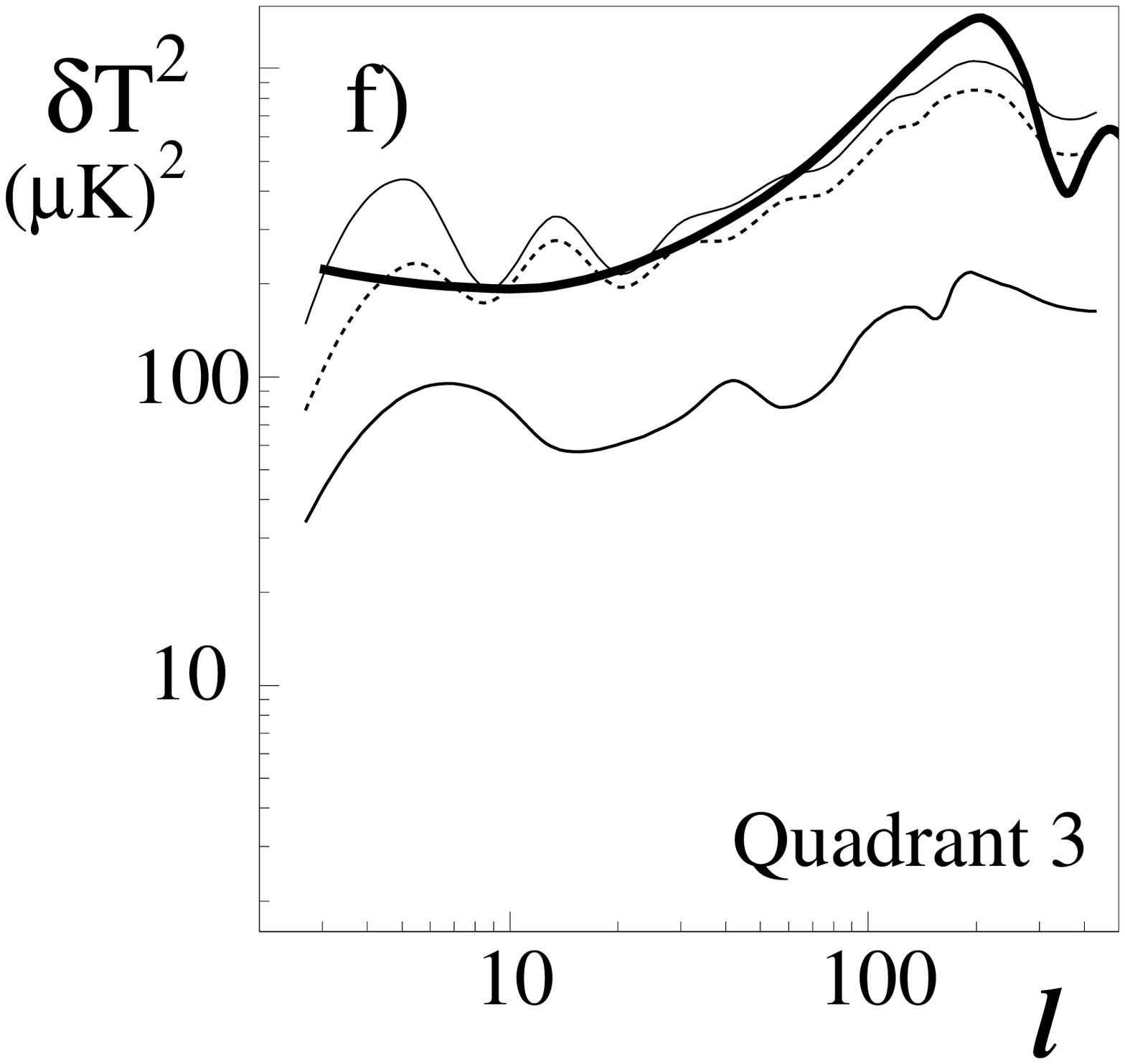}}

\vspace{-1cm}

\caption{(a) Power spectrum, for latitudes below $10^\circ$, for
each Galactic Quadrant in turn.  The
lowest line refers to Quadrant 1, and so on.
Successive lines are displaced
upwards by a decade to help appreciation of their shapes.
(b) As (a) but for latitudes above 10$^\circ$.
(c) The ratio of the power in the other Quadrants to
that in Quadrant 1, for $|b|<10^\circ$.
(d) As (c) but for $|b|>10^\circ$.
Overall Power spectrum and spectra for the two latitude ranges for
(e) Quadrant 1 and (f) for Quadrant 3.  Also shown is the result
\citep{2}
for the whole sky divided by 4 to give the expected
power per quadrant. \label{f2}}
\end{figure}

Figure~\ref{f2}a shows the mean power vs $\ell$ for each Quadrant
for latitudes below 10$^\circ$ and Figure~\ref{f2}b shows results
for latitudes above 10$^\circ$. There is little doubt that the two
$|b|$ regions differ. Figures~\ref{f2}c and \ref{f2}d show
ratios with respect to Quadrant 1.

It is evident that a case can be made for there being big
variations in power, Quadrant to Quadrant, particularly at small
$\ell$ and for $|b|<10^\circ$ these persist to high
$\ell$-values.

Considering, first, $|b|<10^\circ$, and $\ell=10$, the variance
expected has an upper limit $\sim 4.6$, so that Q3/Q1 and Q4/Q1
are on this limit.  Q2/Q1 is below it.  The overall chance
probability for $\ell <
10$ is $\sim$10\%.  At
higher $\ell$, where the variance is smaller, there is an
interesting and significant difference between the Inner Galaxy
(Q1 \& Q4) and the Outer (Q2 and Q3), the power in the Inner is
decidedly higher.  Inner versus Outer is different - for $\ell>50$
- at about the 1\% level.

Turning to $|b|>10^\circ$, the solid angle is bigger
(83\% of the whole) but the excess is smaller.
Interestingly, the significance of 'the effect' is similar, at
least for Q1, Q2 and Q3 with respect to Q1, in the region of
$\ell\sim 5$.  At higher $\ell$, the significance drops,
although above $\ell\sim 100$, the ratios for Q3 and Q4 are
formally significantly above unity.

In terms of the absolute magnitudes, Figures~\ref{f2}e and
\ref{f2}f show comparisons between the power as a function of
$\ell$ for the two Inner and Outer Galaxy Quadrants and that for
the standard CMB model \citep{30}, divided by 4 (ie each for one
quarter of the solid angle).

Some remarks are necessary about these two Figures.  Firstly,
considering the small 'wiggles', these are not significant.
Secondly, the difference between Inner Galaxy (exemplified by Q1)
and Outer Galaxy (Q3) is seen.  Thirdly, the sum of the powers for
each Quadrant is nearer conventional expectation for Q1 than Q3, a
result due to the larger contribution from $|b|<10^\circ$.

\begin{figure}
\centerline{
\includegraphics[width=8cm]{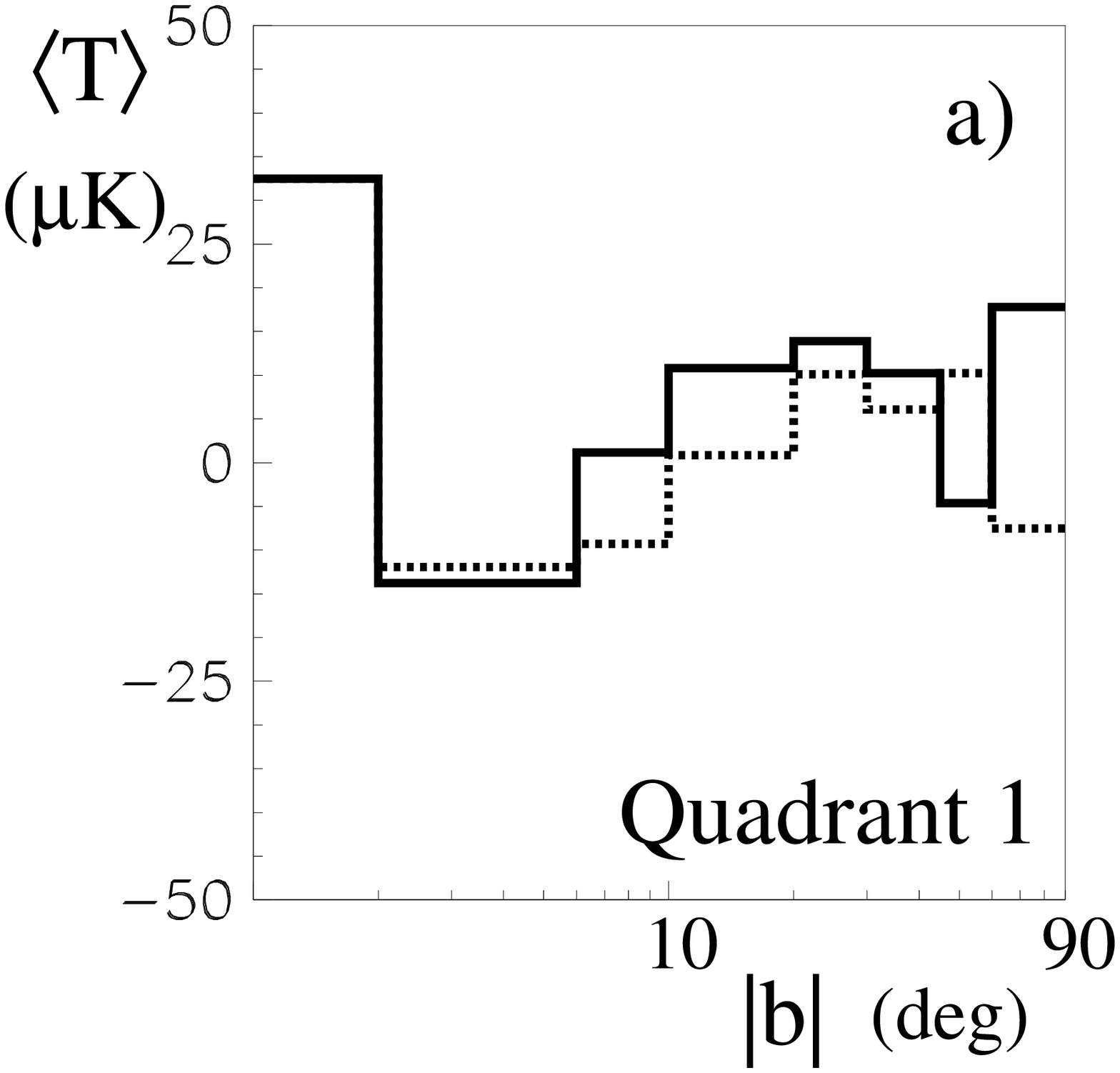}
\includegraphics[width=8cm]{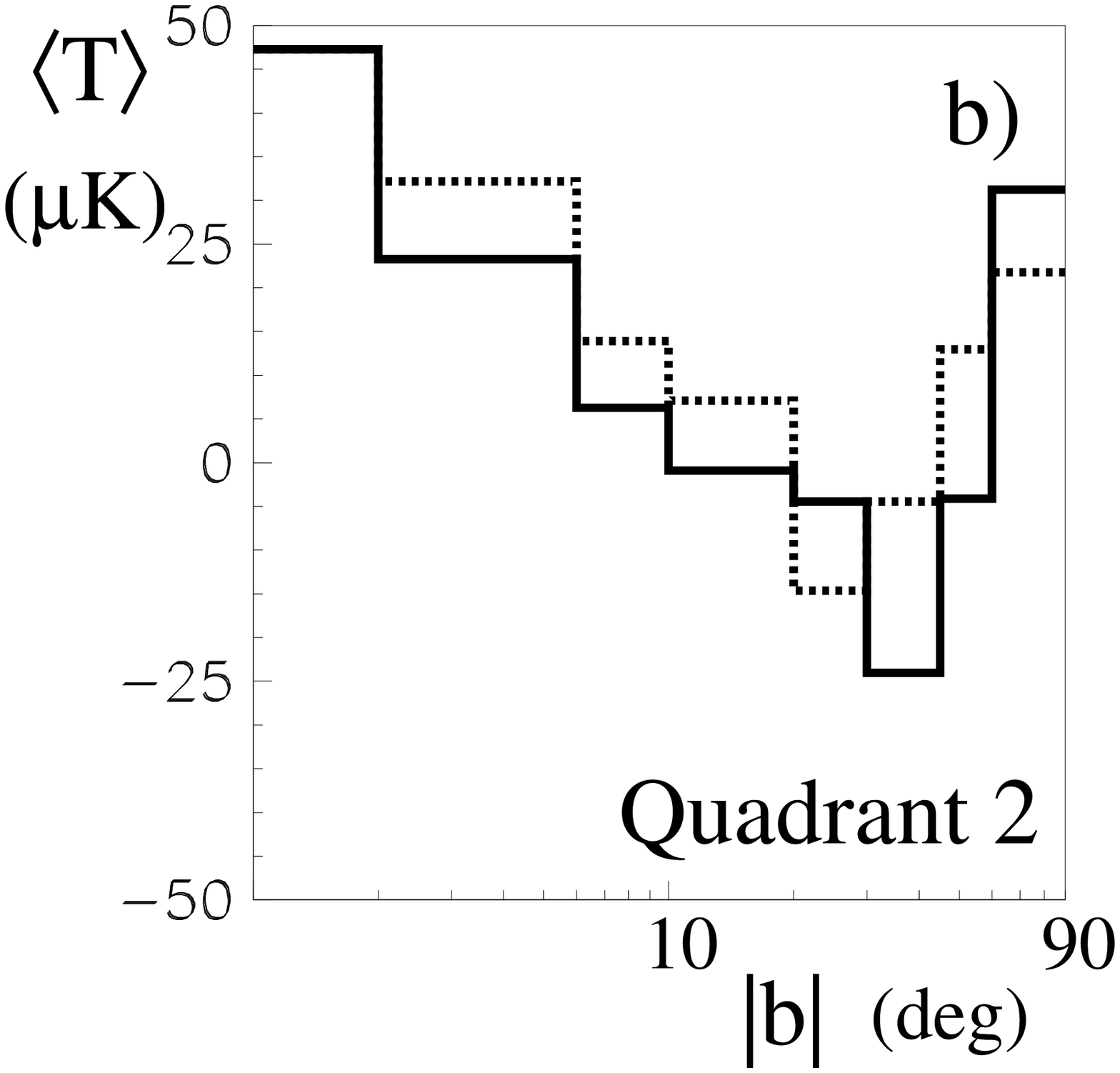}}

\vspace{-1cm}
\centerline{
\includegraphics[width=8cm]{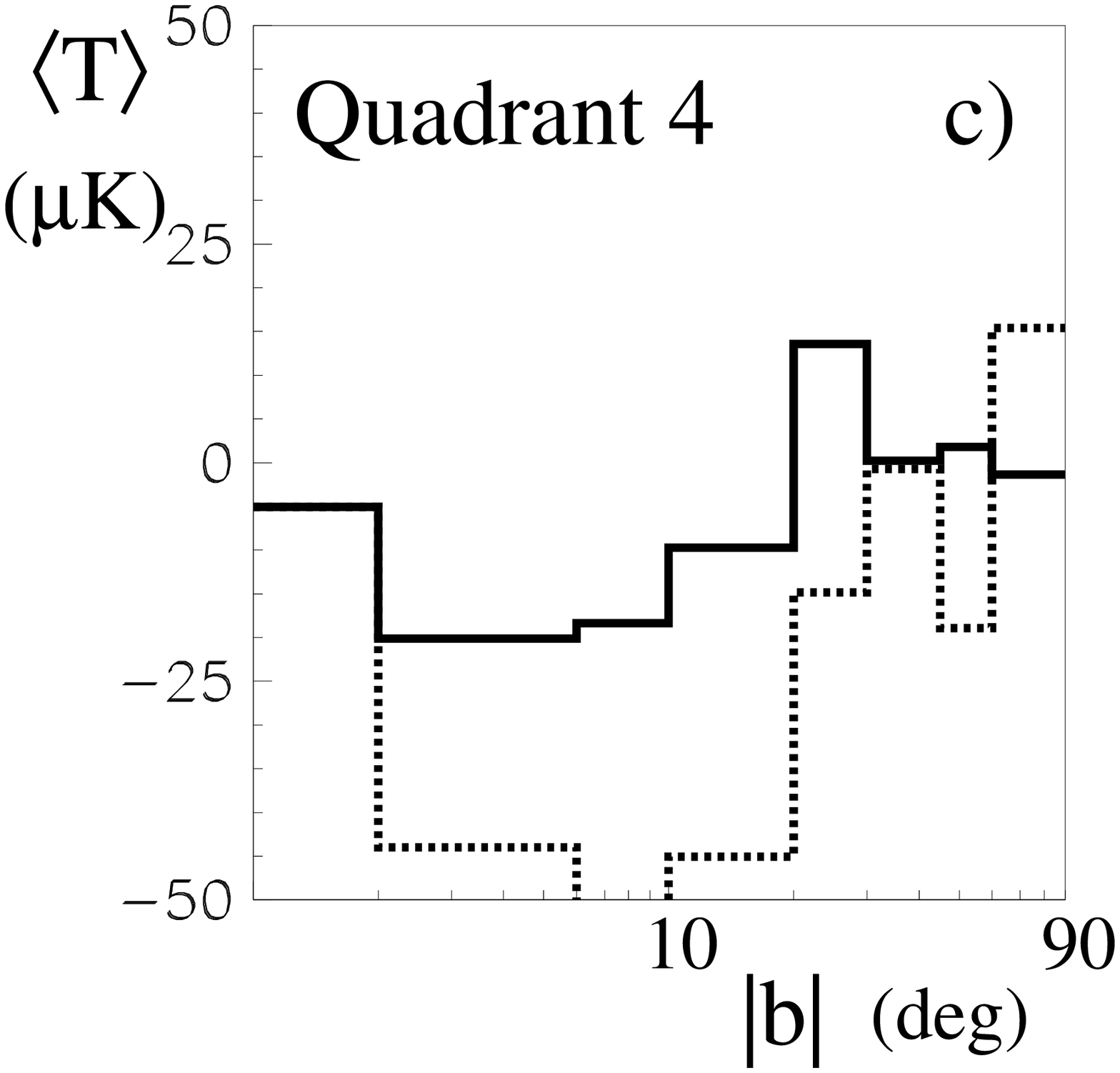}
\includegraphics[width=8cm]{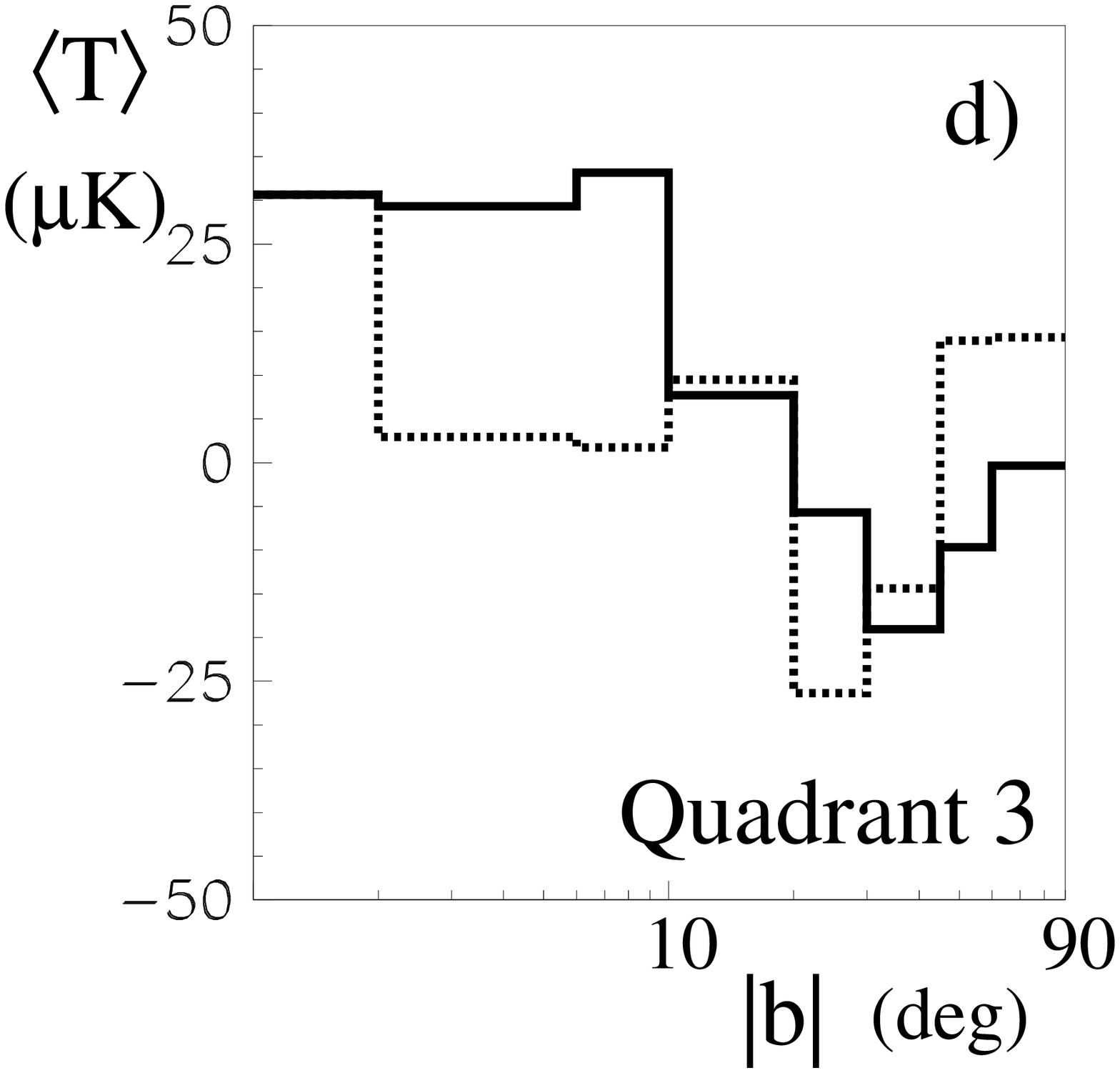}}

\vspace{-1cm}
\centerline{
\includegraphics[width=8cm]{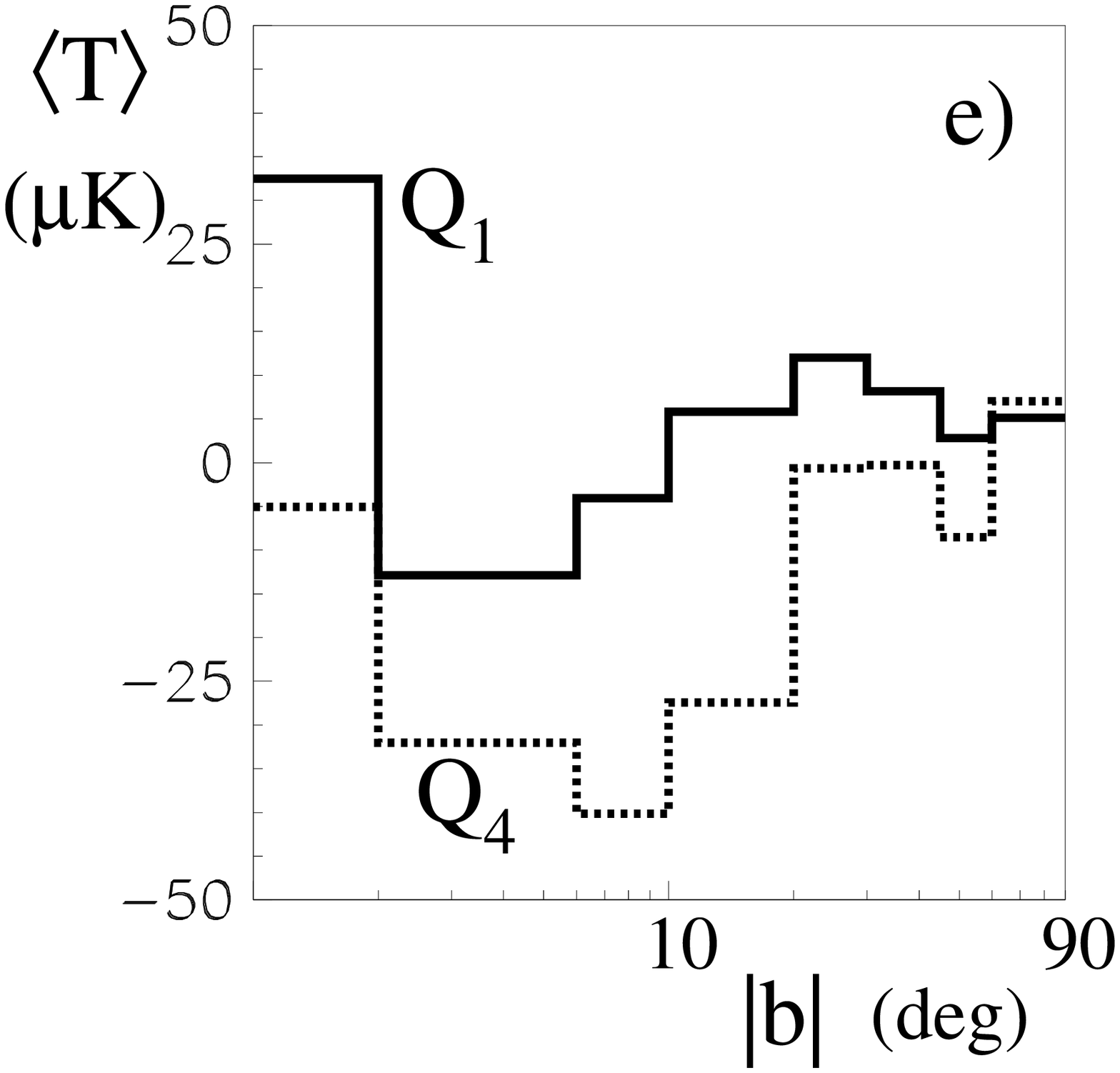}
\includegraphics[width=8cm]{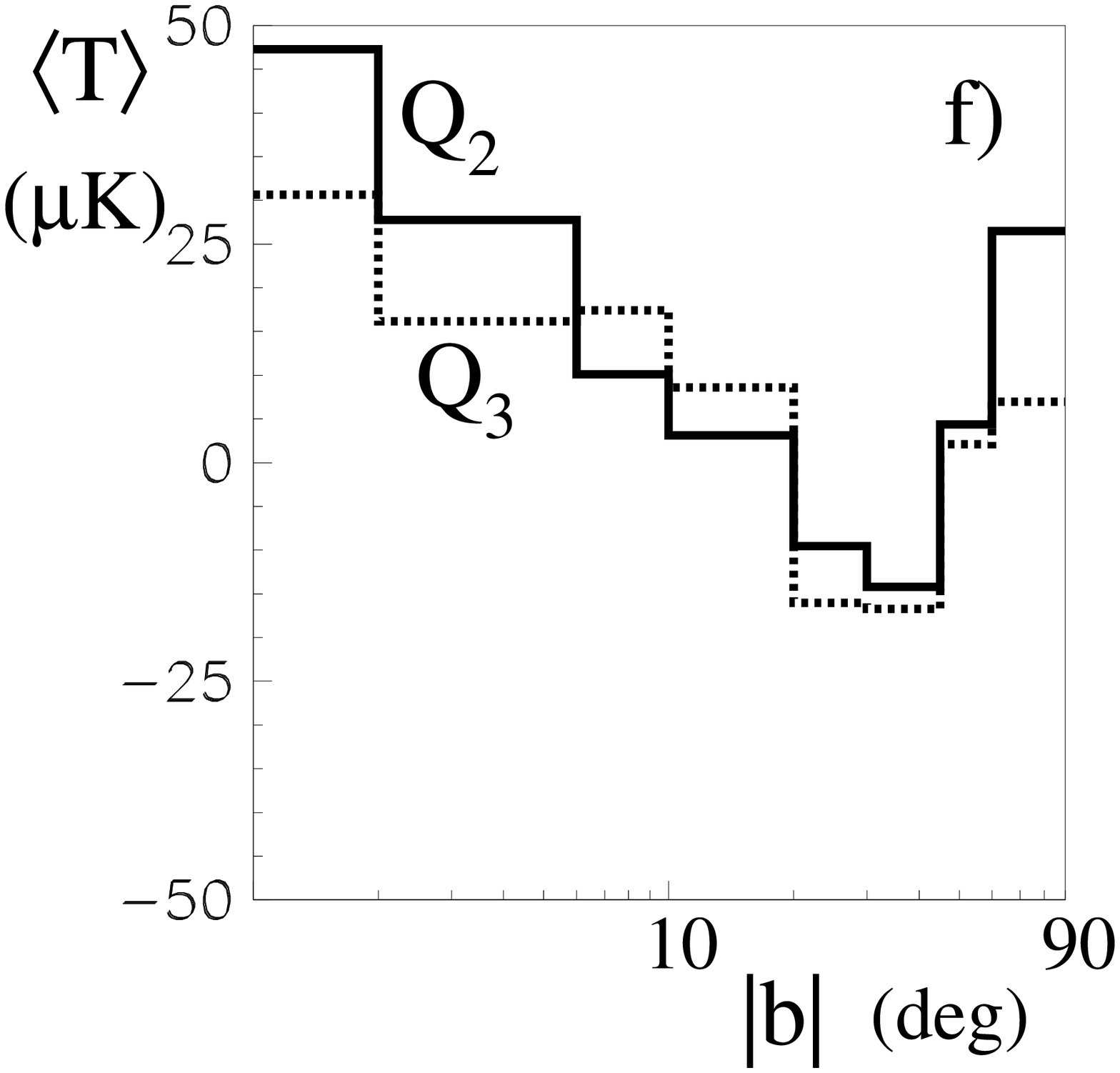}}
\caption{Mean CMB temperature for each Quadrant for North (solid lines)
and South (dashed lines) (a-d), and average values for the whole Quadrants
(e,f) as a function of Glactic latitude.
\label{f3}}
\end{figure}

Further analysis comes from an examination of the overall mean CMB
temperature (as distinct from the power, $\Delta T^2$) as a
function of latitude.  Figure~\ref{f3} shows the results.
Inspection of the figure shows a number of features
\begin{itemize}
\item[i)] There are marked differences Quadrant to Quadrant in
mean temperature. Quantification of a lower limit to the
significance of the variations can be made by assuming that the
variations
within one Quadrant from one latitude band to another are
representative of the errors and deriving the mean displacement
from zero.  Considering $|b|>10^\circ$, where successive latitude
bands are sufficiently independent, we find, taking the 8
'regions' (4 Quadrants, N and S); 4 cases between 0 and 1 standard
deviation, 3 between 1 and 2 and one as high as 5 standard
deviations.  There is thus evidence for significant excursions
from zero.

In the Galactic Plane itself ($|b|<2^\circ$) there is a clear
excess temperature.  The average is 27 $\mu K$ and, if it is valid
to determine the error from the dispersion of the four sets of
data this discrepancy from zero is at the 4 standard deviation
level.  Using the mean error for the other 8 sets of data at
$|b|>10^\circ$ (and five values for each set) the significance
rises to nearly 6 standard deviations. %
The excess is far bigger than one would have expected
from the remarks about cleaning in \cite{3}.

\end{itemize}
Equally important are the symmetries, as follows:
\begin{itemize}
\item[ii)]
Figures \ref{f3}a and \ref{f3}b show clear symmetries
in the $b$ dependence for the two Quadrants in the Inner Galaxy
(Q1 and Q4) and those in the Outer Galaxy (Q2 and Q3).  Turning to
data for the N and S separately Figures \ref{f3}(c) and (d), 
in Quadrants 1 and 2 there is fair symmetry between N and S.
In Quadrants 3 and 4 there is some symmetry (but less than for
Quadrants 1 and 2) for $|b|>10^\circ$ (it is relevant to point
out that the giant 'hole' in the CMB, at $\ell \sim 300^\circ$ to
360$^\circ$, $b \sim 0^\circ$ to $-30^\circ$, is responsible
for the very low mean temperature in Quadrant 4 for $b$:
$-2^\circ$ to $-20^\circ$.
%
The symmetry can be quantified by
comparing the separation of the two values, at each $|b|$, with
$\surd2$ times the modules of the distribution of the individual
values about zero.  The result for the ratio of the median values,
(equivalent to a number of standard deviations) is, for Q1, Q2, Q3 and
Q4; 2.1, 2.1, 1.4 and 1.1.  The symmetry is seen to be reasonably
significant for Q1 and Q2.  Taken overall, N-S symmetry is present
at the 2.4 sigma level.

\item[iii)] There is no systematic excess temperature in the South
as there is in the power spectrum at low $\ell$ (Figure~\ref{f2}),
at least averaged over all $b$ (but see later).

\item[iv)] In the Outer Galaxy (Quadrant 2 and 3) there is some
evidence for a minimum in the range $|b| = 30^\circ -40^\circ$, ie
a symmetrical feature. 
 \item[v)] In the Inner Galaxy
(Quadrants 1 and 4) there is a dip in the range 2$^\circ -
20^\circ$ in all four cases, again a symmetrical feature.

\item[vi)] In every case the mean temperature for $|b|<2^\circ$
is higher than that for $|b| = 2^\circ - 6^\circ$ and higher
than that averaged over the $|b|= 2^\circ - 20^\circ$ range.

\item[vii)] There is a fall in $\langle\Delta T\rangle_{\rm CMB}$ with increasing
$|b|$ over the range $|b|=6^\circ-45^\circ$, after which
there is an increase.

\item[viii)] Figures~\ref{f3}(e) and \ref{f3}(f) show good symmetry
between the two Inner Quadrants (1 and 4), and 
between the two Outer Quadrants (2 and 3).

\end{itemize}
The conclusions to be drawn at this stage from the results on
$\langle\Delta T\rangle$ are:
\begin{itemize}
\item[i)] There is a clear excess at $|b|<2^\circ$ averaged over
all Quadrants and thus there is certainly some foreground left in
this range.
\item[ii)]Whilst there are clear Galactic features these are not
of simple cosec $b$ form over the whole range of $|b|$;
nevertheless there is a reduction of temperature from
$|b|=6^\circ$ to $45^\circ$, the region with most of the solid
angle of the sky.
\item[iii)] There are interesting symmetries
about the Galactic Plane.
\end{itemize}
Comparison with the other results on 'Gaussianity', etc will be
made later.

The results of the mean temperature are not necessarily
inconsistent with those for the power spectrum (Figure~\ref{f2})
because the mean temperature relates to a form of integral over
all $\ell$-values, whereas the power spectrum refers to the square
of the temperature and is, of course, $\ell$-dependent. This
aspect can best be studied by examining plots similar to Figure
\ref{f3} for the power (per unit solid angle) for different
$\ell$-values. We remember that the power spectrum represents the
power per unit logarithmic interval in $\ell$.

\begin{figure}
 \centerline{
\includegraphics[width=9cm]{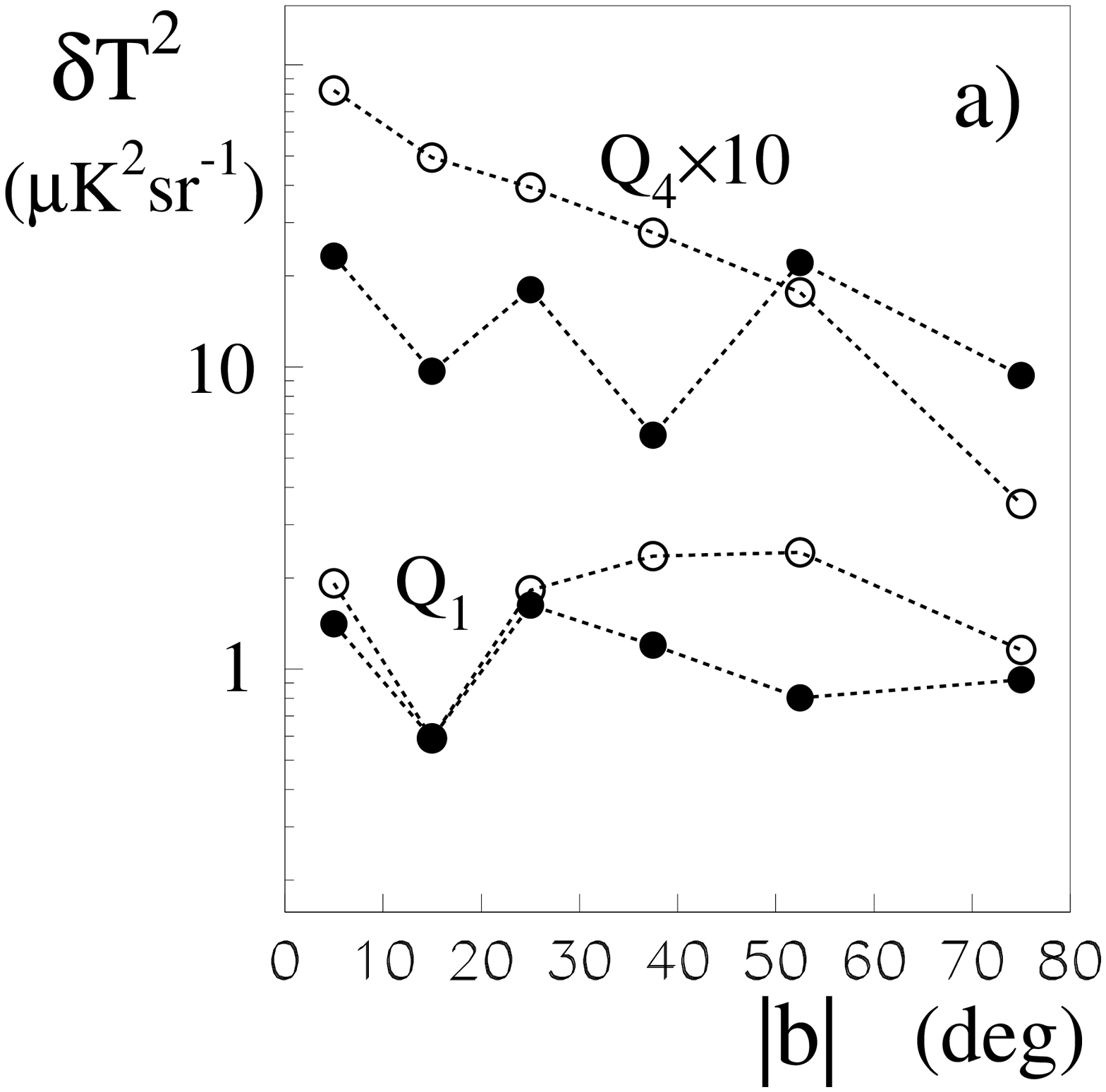}
\includegraphics[width=9cm]{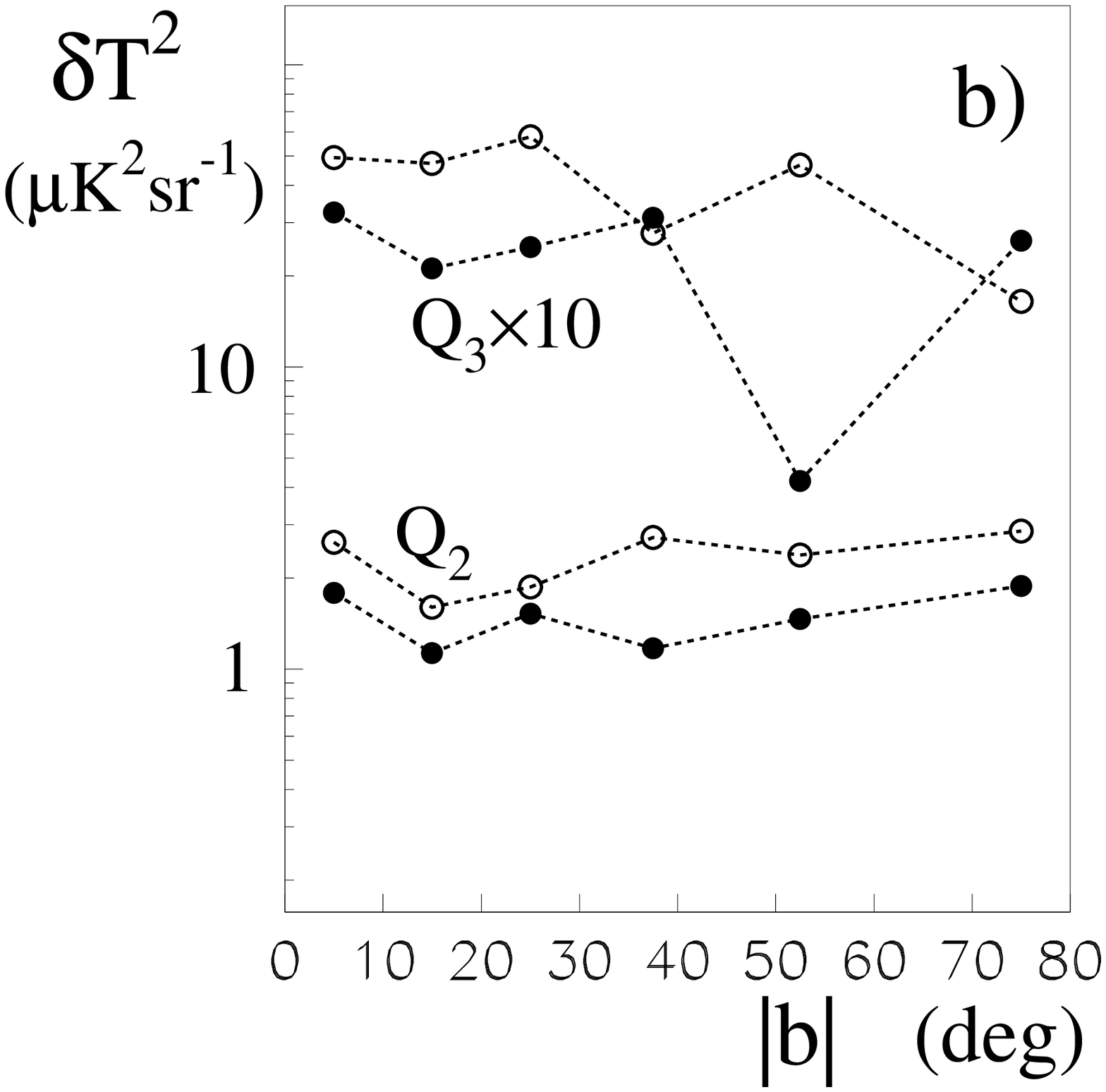}}
 \centerline{
\includegraphics[width=9cm]{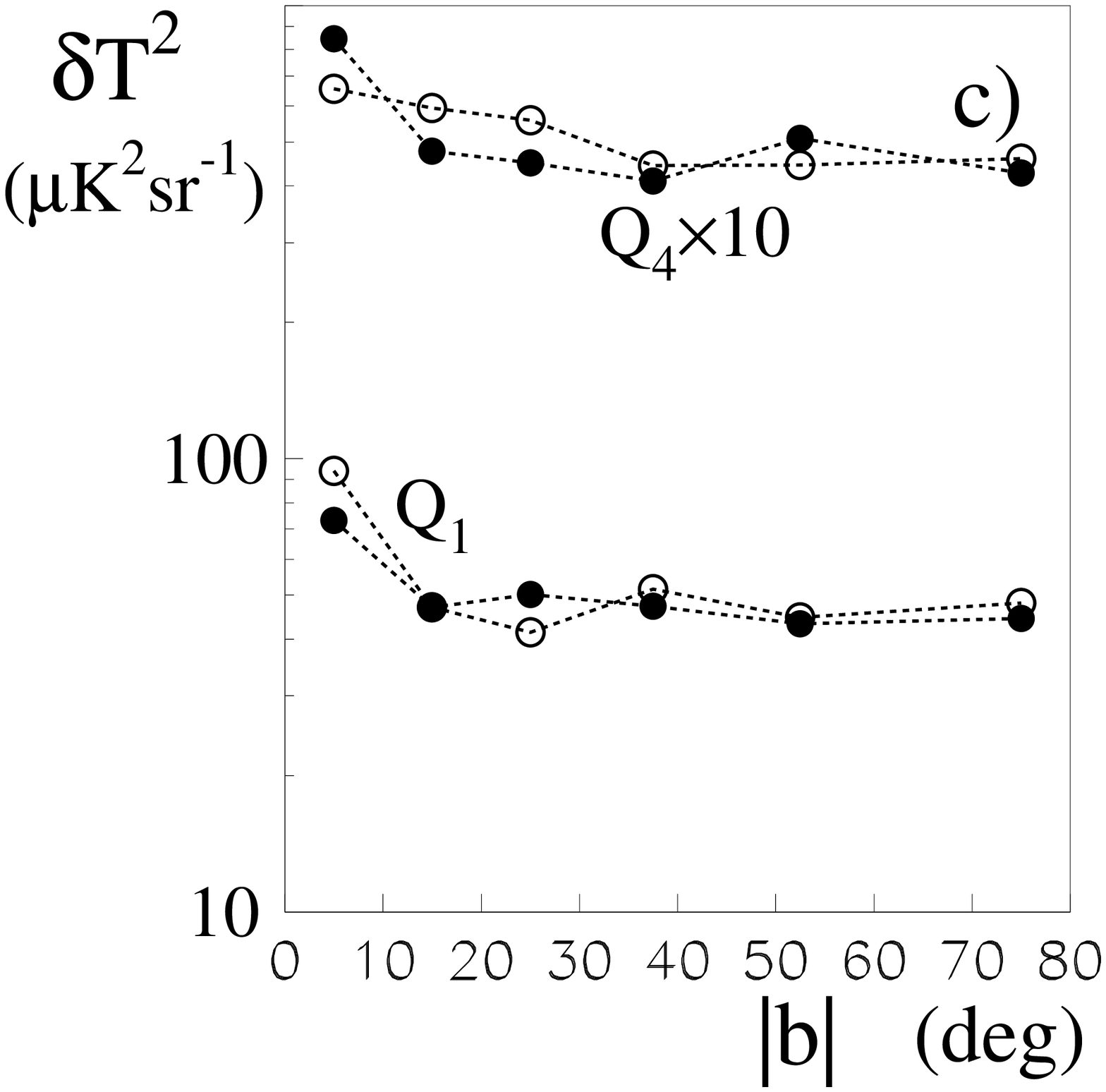}
\includegraphics[width=9cm]{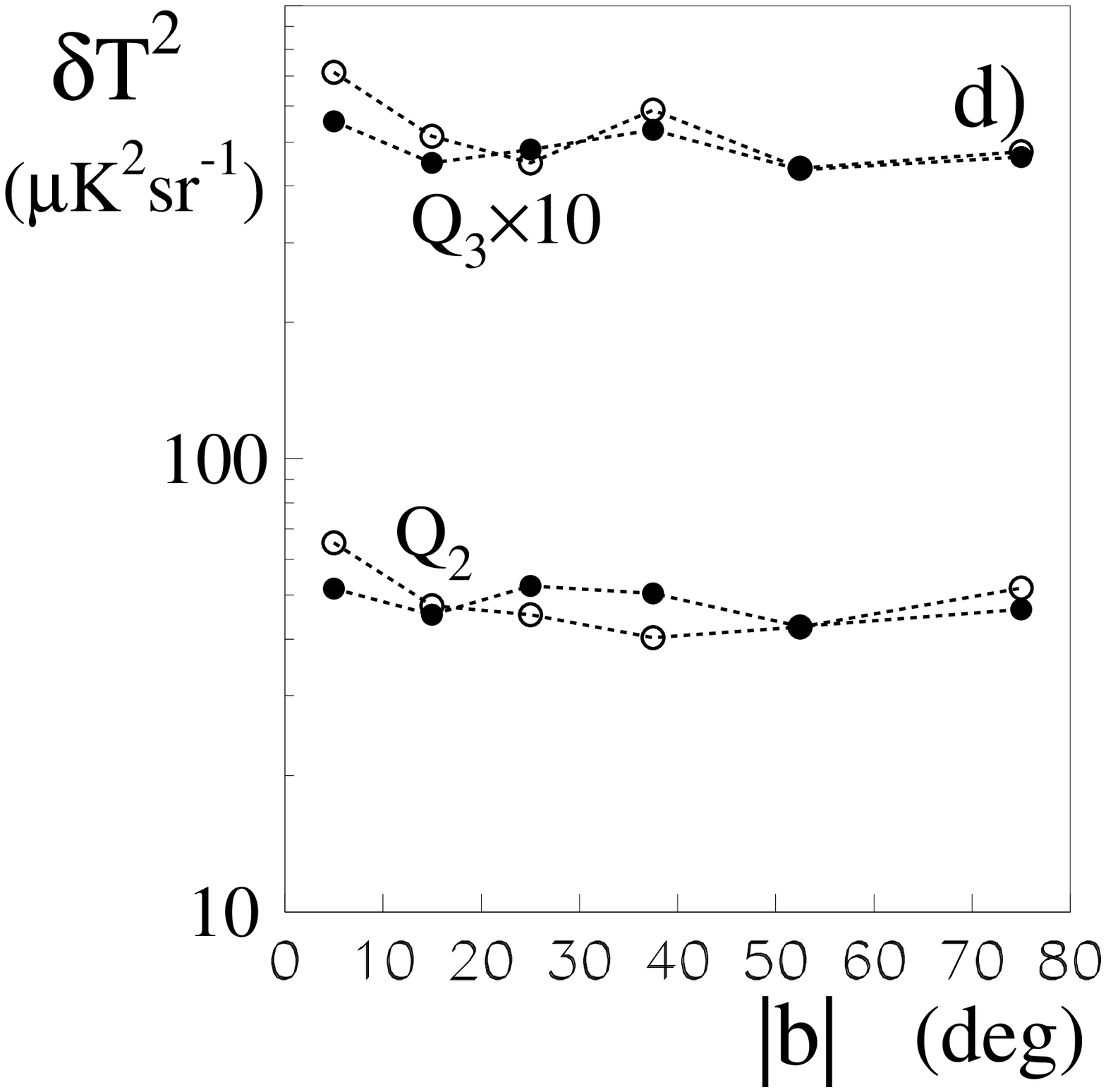}  }
\caption{(a) and (b)
Power at $\ell=10$ for the different latitudes and Quadrants.
(c) and (d) as (a) and (b) but for $\ell=100$.
The power is expressed per unit solid angle.
The open circles relate to the S-hemisphere
and the filled circles relate to the North.  It is evident that
there are
significant variations.
 \label{f4}}
\end{figure}

Figure \ref{f4} shows the results for two values of $\ell$, $\ell
= 10$ and $\ell = 100$.  The main feature here of relevance to
symmetry are:
\begin{itemize}
\item[i)] Again, there is an excess nearest the Galactic Plane -
in this case for $|b| < 10^\circ$. The significance has been
examined as for mean temperature, deriving the mean value for the
other latitudes, and its error, and finding the difference between
the value for $|b|<10^\circ$ and this mean.  There are 8 sets of
data and the result is that for $\ell=10$ the average difference
is 2.2 standard deviations - ie over 5
 standard deviations overall. For
$\ell$=100, the significance is even bigger, amounting to 10
standard deviations. For $\ell=100$, where the interdependence of
different $|b|$-bands is least, there is a strong difference
between Inner and Outer Galaxy.  In the range $|b|<10^\circ$, the
excess in the Inner Galaxy is 2.6 times that in the Outer.  The
significance is 3.2 standard deviations. For $|b|>10^\circ$ the
excess changes to a deficit.  The ratio is $0.72 \pm 0.05$, ie 5.6
standard deviations from unity. There is clearly no evidence for a
constancy of power.

\item[ii)] In every one of the 16 cases (for both $\ell$-values)
the power in the $|b| = 10^\circ - 20^\circ$ is lower than
would be expected on the basis of the adjacent points; thus a
Galactic symmetry.

\item[iii)] For $\ell$=10, in three out of four Quadrants there is
a slow fall of power with increasing latitude beyond $|b|=
20^\circ$. A plot of $\sqrt{\langle\Delta T^{2}\rangle}$ vs cosec
$|b|$ in this range yields a slope of $0.3 + 0.04/-0.15$ $(\mu
K^{2}sr^{-1})^{1/2}$ for $|b| > 20^\circ$.  The mean S/N ratio in
this region can be derived from Figure~\ref{f1}d.  It is 1.3 (the
square root of the derived value, which related to $\Delta
T^{2}$), to be compared with $\sqrt{1.7} = 1.3$, for all
longitudes at $\ell = 1$, (Figure~\ref{f1}c). An important result
is that the mean slope for the South is greater than that for the
North, although this result comes mainly from Quadrant 4. 
None of the features described above would be expected
for 'cosmic variance'. 

\item[iv)] At $\ell = 100$, the effects are too small, in
comparison with the errors, for a pattern to be discerned with
respect to $|b|$ dependence and N-S ratio.
\end{itemize}

The conclusions here are that, again, there is residual foreground
near the Galactic Plane, the range $|b| = 10^\circ - 20^\circ$
appears to have been overcorrected and that, at low $\ell$, there
is an all-latitude foreground.  Furthermore, there is evidence for
a cosec $|b|$ dependence of the root mean square power (at the
2$\sigma$ level).

Turning to the N-S asymmetry, we note that the S/N ratio for
power, is, for $\ell$=10, and increasing $|b|$: 2.11, 2.49, 1.79,
1.85, 2.25 and 0.95.  leaving out the (anomalous) Quadrant 4, the
S/N ratios are: 2.39, 2.69, 1.53, 2.83, 1.50 and 0.72.

The S/N ratio therefore dips towards unity at $|b| \sim -60^\circ$
in the second case. Inspection of the 408 MHz results \citep{8},
shows that there is a reduction in ratio starting at $|b|\sim
40^\circ$ reaching a minimum at $|b|\sim 60^\circ$.  The
individual frequency bands for WMAP (uncleaned) show a ratio of
$\sim$1.37(+0.13-0.04) at $|b|$=30$^\circ$ (and little variation
at lower $b$), 1.21 ($+0.04-0.02$)
at $|b|$= 40$^\circ$ and 1.06
(+0.04-0.02)at $|b|$=60$^\circ$, the limits encompassing the whole
range of frequencies.  It is interesting to note that there is no
systematic change to the S/N ratio with frequency; but both in the
408 MHz and the individual WMAP frequencies there is a reduction
in ratio at mid-latitudes; 40$^\circ$ at 408 MHz and
$\sim$40$^\circ$ in the individual WMAP frequency $\Delta
T$-values, to be compared with 45$^\circ$ - 60$^\circ$ in the
$\ell$= 10 WMAP power.  Once again, then, there is evidence of
insufficient cleaning.

An important aspect is a comparison of the two Quadrants in the
outer Galaxy, Q2 and Q3 in both $\langle T \rangle$ and power; we would expect
greater activity in Q2 in comparison with Q3 because of the
greater number of SNR, HII regions, etc there.

Inspection of Figure~\ref{f3} shows that the temperature in Q2 is
higher than that in Q3 for every latitude bin, except for the
(anomalous) region $|b|: \sim 10^\circ \div 20^\circ$
and the previous bin.
The excess (Q2-Q3) in $\Delta T$
is successively, for $|b|=0^\circ -2^\circ$, $2^\circ - 6^\circ$,
, $>60^\circ$:
%
+17, +12, -7, -5, 7, 3, 3 and 19 $\mu K$.  An excess is thus well
founded.  Insofar as much of $\Delta T$ comes from high
$\ell$-values (and thus small angular sizes), there should be
little cross-talk between the two Quadrants.

Surprisingly, perhaps, the same feature (Q2$>$Q3) does not pertain
for the power per steradian (Figure~\ref{f4}). For $\ell=10$, only
for the highest latitudes ($|b|>60^\circ$) is Q2$>$Q3 -- and here we
would expect the nearest approach to equality.
For
$|b|>10^\circ$ the ratio for Q2/Q3 is 0.56, for
$|b|=10^\circ-20^\circ$ it is 0.43 and the average for
$|b|>20^\circ$ is 0.69.  At $\ell = 100$, a similar situation
pertains: for $|b|>10^\circ$, Q2/Q$3 = 0.93$, for
$|b|>10^\circ-20^\circ$, Q2/Q$3 = 0.97$ and for $|b|>20^\circ$
Q2/Q$3 = 0.94$.  Again, in all latitude bins, Q2/Q3$>$1 with the
exception of $|b|> 60^\circ$ where Q2/Q$3 = 1$.

Thus, there is a clear difference in the power per steradian for
Q2 and Q3 and one that is systematic over all latitudes (see
Figure~\ref{f2}).  At first sight it is of the opposite sense to
that expected but the answer may lie in the fact that the radial
gradient of the CR intensity is (surprisingly) smaller for Q3 in
comparison with that for Q2 \citep{25}.

Turning to the S/N ratios for Q2 and Q3, overall (over all $b$)
these are: Q2, 1.63 (1.04); Q3, 1.79 (1.09).  The first value
related to $\ell=10$ and the second to $\ell =100$.

The corresponding values for $b>20^\circ$ are: 1.69 (0.96) and
1.72 (1.02).  The ratio is thus rather stable with respect to the
limit on $|b|$. A different situation pertains for $\langle T
\rangle$, where the differences (S-N) are, for Q2, $+33~\mu K$
($+16$) and for Q3 $- 6~\mu K$ ($+25~\mu K$) the first value is
for $|b|>2^\circ$ and that in brackets is for $|b|>20^\circ$.  The
fact that a S-excess is present for $|b|>20^\circ$ in both power
and $\langle T \rangle$ is reassuring, 
 as is the similar magnitude for both Quadrants.

\subsection{Comparison with the original WMAP and 408 MHz data }
\label{iiD}

A further aspect of the 408 MHz data is the presence of 'fine
structure' in the cosec $|b|$-plots' and also in the separate
plots of $\Delta T$ vs cosec $|b|$.  In every one of the six plots
\citep{7}
 there are consistent features,
in order of diminishing magnitude:

 \begin{itemize}
 \item[-]  Minima near cosec $|b|$ = 2.0 (ie $|b| = 30^\circ$)
 \item[-]  Minima at cosec $|b|$ = 3.15 ($|b| = 18.5^\circ$)
 \item [-] Minima at cosec $|b|$ = 1.15 ($|b| = 60.4^\circ$), all
 for the North.
 \end{itemize}
 For the South there are:
 \begin{itemize}
 \item[-]  Minima at cosec $|b|$ = 1.15 ($|b| = 60.4^\circ$)
 \end{itemize}
 And, for the five WMAP frequencies alone, there are:

 \begin{itemize}
 \item[-] minima at cosec $|b|$ = 2.15.
 \end{itemize}

 In the North alone at the smallest angle (cosec $|b| = 4$, ie
 $|b| = 14.5^\circ$) there is an excess over the 'best straight line'
 growing systematically with frequency, reaching 17\% at 93~GHz.  There is no equivalent in the
 South.

 Such fine structure, even when the data are integrated over all
 longitudes, lead us to believe that significant variations,
 correlated in frequency, undoubtedly exist for smaller longitude
 bins - as we claim.  It is very doubtful if such effects would
 be subtracted out in the WMAP cleaning procedure \citep{8,17}
 - in  which the weights vary considerably from one frequency to
 another. (They are, in order of increasing frequency, $23 - 93$~GHz, $+0.109$,
 $-0.684$, $-0.096$, $+1.921$ and $-0.250$).  Furthermore, the weights are
 the same for N and S, whereas the dependence of slope of
 $\Delta T$ vs cosec $|b|$ is different for N and S.  Specifically,
 the slopes are, for $23-93$~GHz:
 $(\delta\Delta T/\delta ({\rm cosec}|b|))_{\rm S/N}
 =1.50$, 1.47, 1.67, 1.61 and
 1.50.  That for 408~MHz is lower: 1.10 although, because the
 dependence of T on cosec $|b|$ is less linear, there is an
 appreciable error.

\subsection{The Situation at High Latitudes} \label{iiE}

It is instructive to examine the situation beyond $|b|=45^\circ$
in some detail because this region had been expected to be far
enough from the Galactic Plane to be completely free of foreground
\citep{30}.

Other work \citep{12}
, mentioned already,
has suggested that the axis of symmetry of a large scale
modulation of the CMB signal is in fact, not along the Galactic
N-S axis but nearer the axis through the Ecliptic Poles. These
results will be discussed later in \citep{SF}; here, we examine
this aspect from a different standpoint.  Ideally, with a large
foreground contribution, we might expect the 'N-S axis' to be
defined by the lowest signals in the N and S hemispheres. However,
even in the case of conventional Galactic radiation signatures (eg
21 cm line of the HI column density, synchrotron radiation, etc)
the minima are away from the Galactic Poles. Furthermore, as we
have noted, some of our CMB parameters \textit{increase} with
latitude above $|b|$ = 45$^\circ$.  Thus, we determine the
directions of both minima and maxima for each hemisphere.  The
parameters studied comprise those from the CMB analysis already
given, viz $\Delta T_{\rm CMB}$, W($\ell$=10) and the correlation
coefficient for CMB temperature with respect to the CR intensity
(to be described). In addition, we include directions for low (and
high) spectral exponents for CR protons and electrons
\citep{19,20}, making 5 parameters in all. With the exception of p
and e, the directions are imprecise in that they are derived from
the Quadrant - wide regions and $|b|$ ranges $45^\circ - 60^\circ$
and $>60^\circ$; poor statistics preclude a more precise study.

The results for the range of $l$ and $b$, and the mean values,
are as follows.

Maxima
\begin{itemize}
\item [N:] $l$-range: 120$^\circ$, mean $l$:90$^\circ$
$\pm 30^\circ$

\hspace{3mm}$b$-range: $20^\circ$, mean $b$: +50$^\circ$ $\pm 5^\circ$
\item [S:] $l$-range: 90$^\circ$, mean $l$:230$^\circ$ $\pm$
25$^\circ$

\hspace{3mm}$b$-range: 15$^\circ$, mean $b$: -50$^\circ$ $\pm$5$^\circ$
\end{itemize}
Minima
\begin{itemize}
\item [N:] $l$-range: 180$^\circ$, mean $l$: 200 $^\circ$
$\pm$ 50$^\circ$

\hspace{3mm}$b$-range: 20$^\circ$, mean $b$: $+55^\circ \pm 5^\circ$

\item [S:] $l$-range: $130^\circ$, mean $l$: $20^\circ \pm
35^\circ$

\hspace{3mm}$b$-range: $20^\circ$, mean $b$: $-55^\circ$
$\pm 5^\circ$
\end{itemize}

Inspection of the values shows that indeed the means of both
maximum and minimum values do not centre on $b=\pm 90^\circ$.

The analysis is such that a uniform (ie random) set of directions
would have a range of $l$ - 360$^\circ$ and a mean with an
error in $\ell$ of 
$\pm$90$^\circ$, ie bigger than observed.

The result is that the directions are consistent with conformity
as expected if foreground effects are significant at the high
latitudes - to the extent of about 2-3 sigma in the North and
somewhat greater in the South. The small errors for the $b$-values
are perhaps illusory because of the small overall range of $b$
considered: '$\Delta b=\pm 11.25^\circ$ (from the centre of one
bin to the centre of the other).

Interestingly, the South, with its higher power, 
 Figure~\ref{f1},
is more significant.  Comparison with the other result \citep{12}
is given later.

At this stage, we can claim some support for the hypothesis that
the foreground effects contribute to 'high' latitudes although to
use the statistical weights just referred to is unwise since we
run the risk of double counting, having already used the e and p
result and some aspects of $\Delta T$ - and power - variations.

 \subsection {Comparison with other studies of asymmetries in the cleaned
maps}\label{iiF}

Mention has already been made of the 'Gaussianity' differences.
Work on comparing the statistic for each Galactic hemisphere
(\cite{32}
, b)
 has shown significant differences from expectation
for the North but not for the South.  The nearest we come to an
effect of this sort is the variations from bin to bin in latitude
of the power per sr at low $\ell$ ($\ell\sim 10$); Figure~\ref{f4}.
In two of the Quadrants, 3 and 4, the variation is bigger in the
North than in the South.  The conclusion that there is 'an
unexpected power asymmetry between the N and S hemispheres in the
WMAP data' is certainly supported by our work.  The authors of the
Gaussianity work referred to above have gone on to
use both the power spectrum and N-point correlation functions to
argue that the axis of maximum asymmetry is close to the ecliptic
axis. A somewhat similar result has come from an analysis of 'the
local curvature of WMAP data' \citep{12}
 although here anomalies
are found for both hemispheres, at least for smoothing and the
maximum asymmetry is between the ecliptic axis and the Galactic
Pole. Interestingly, it is demonstrated that for $5^\circ -
10^\circ$ smoothing the Southern hemisphere yields results too
close to the Gaussian expected values.  Inspection of the map for
'lake counts' ('lakes' are regions where the differential of
temperature vs position is changing from negative to positive)
shows maximum fluctuations which correlate reasonably with the
'hot regions' in Figure \ref{f3}.  In fact, inspection of the map
presented in \cite{12}
 shows that the centroids of the N and S
hemisphere regions having unusually high fluctuation values are
somewhat away from the celestial poles.

Specifically, they are at
$l$, $b$:90$^\circ$, +50$^\circ$ and 270$^\circ$, -45$^\circ$.

These values are close to our estimates for the mean positions of
the maxima 

\begin{itemize}
\item[-] $l,b$: $90^\circ \pm 30^\circ$, $+50^\circ \pm 5^\circ$ and
\item[-] $l,b$: $230^\circ \pm 25^\circ$, $-50^\circ \pm 5^\circ$.
\end{itemize}

 The near agreement can be regarded as a confirmatory check on our
 analysis - and thus an agreement - to some extent - with our
 contentions.

\subsection{Power spectra for 'red' and 'blue'
maps}\label{iiG}

\begin{figure}
\centerline{
\includegraphics[width=9cm]{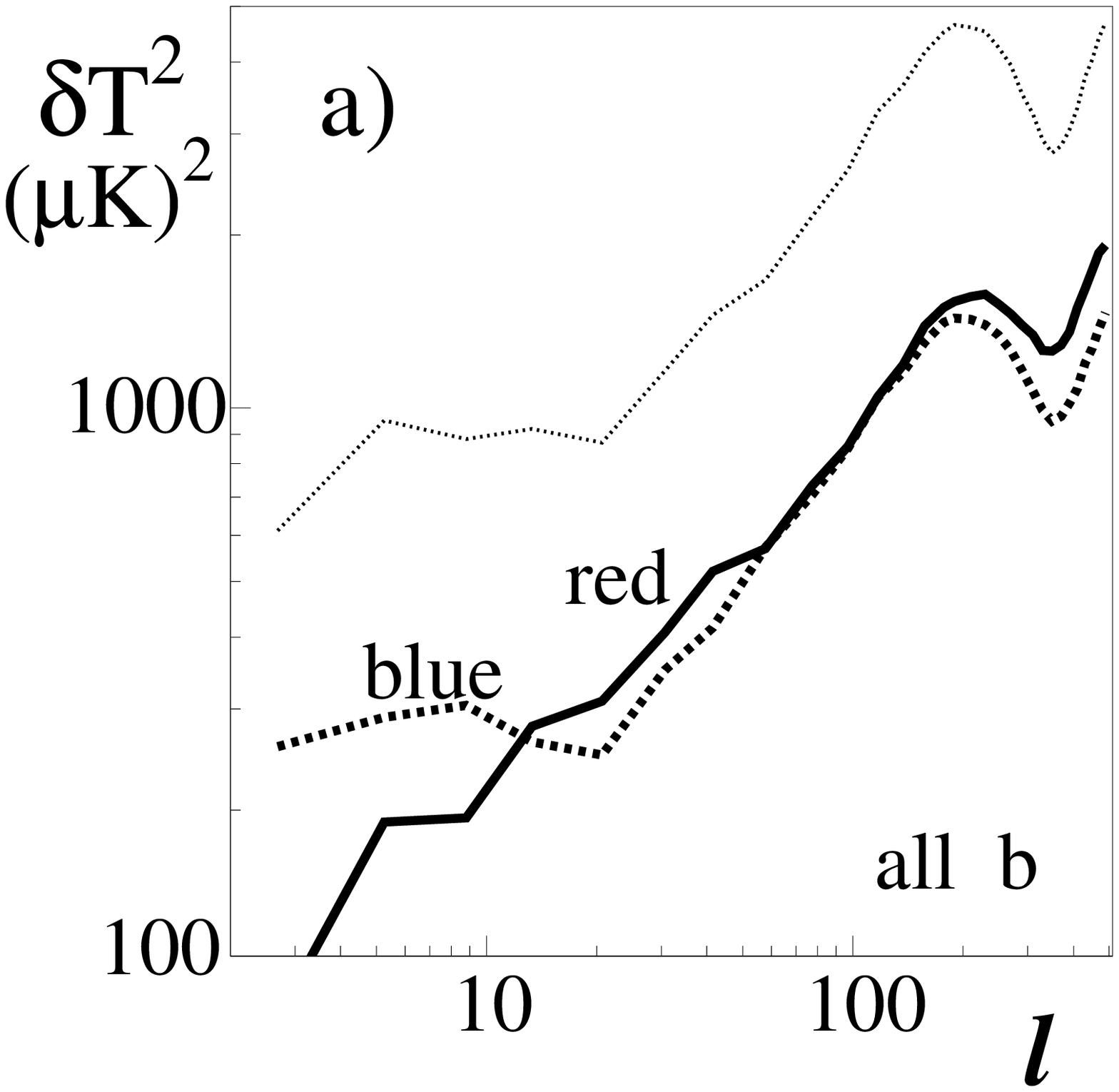}
\includegraphics[width=9cm]{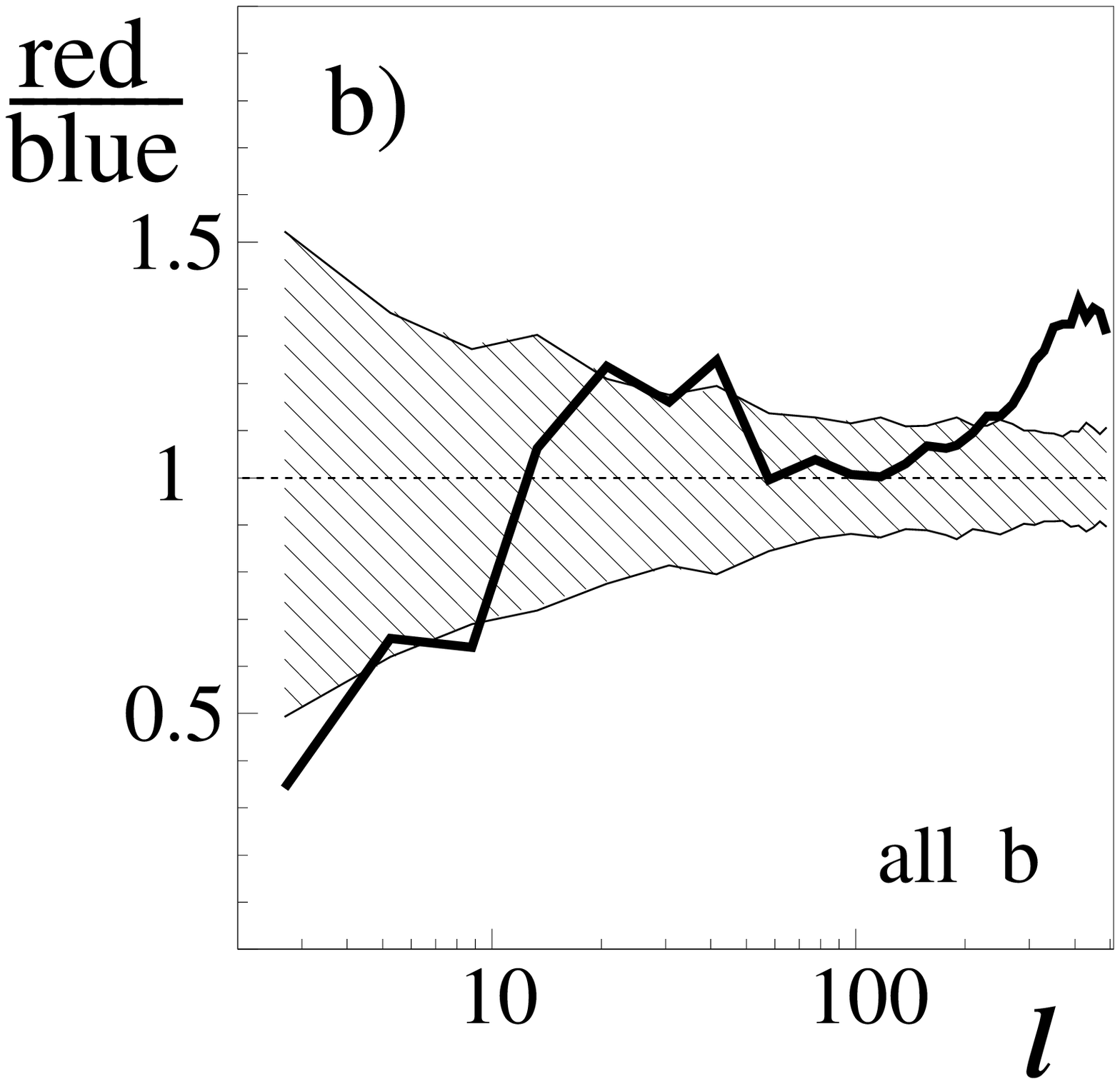}
}
\centerline{
\includegraphics[width=9cm]{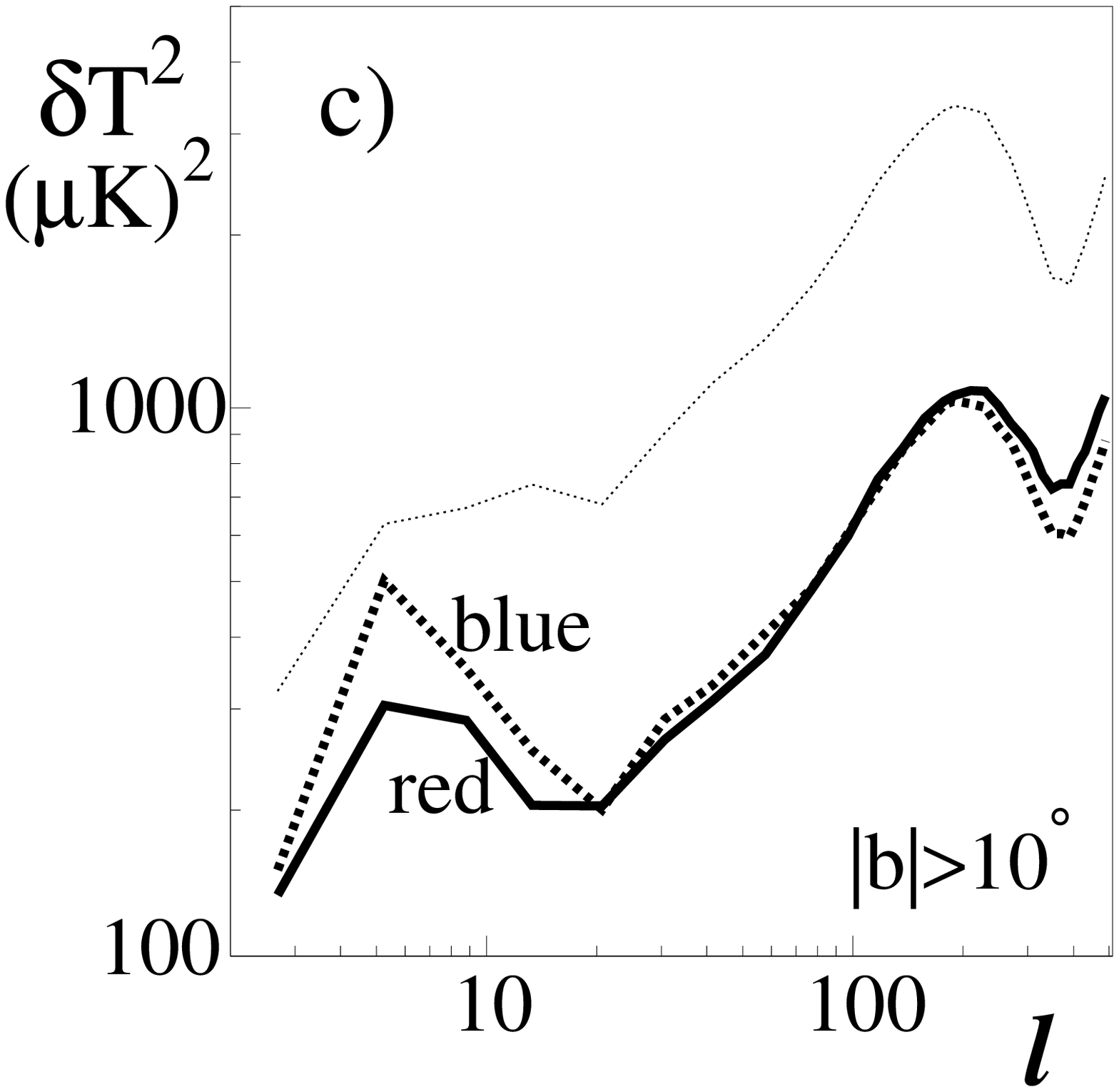}
\includegraphics[width=9cm]{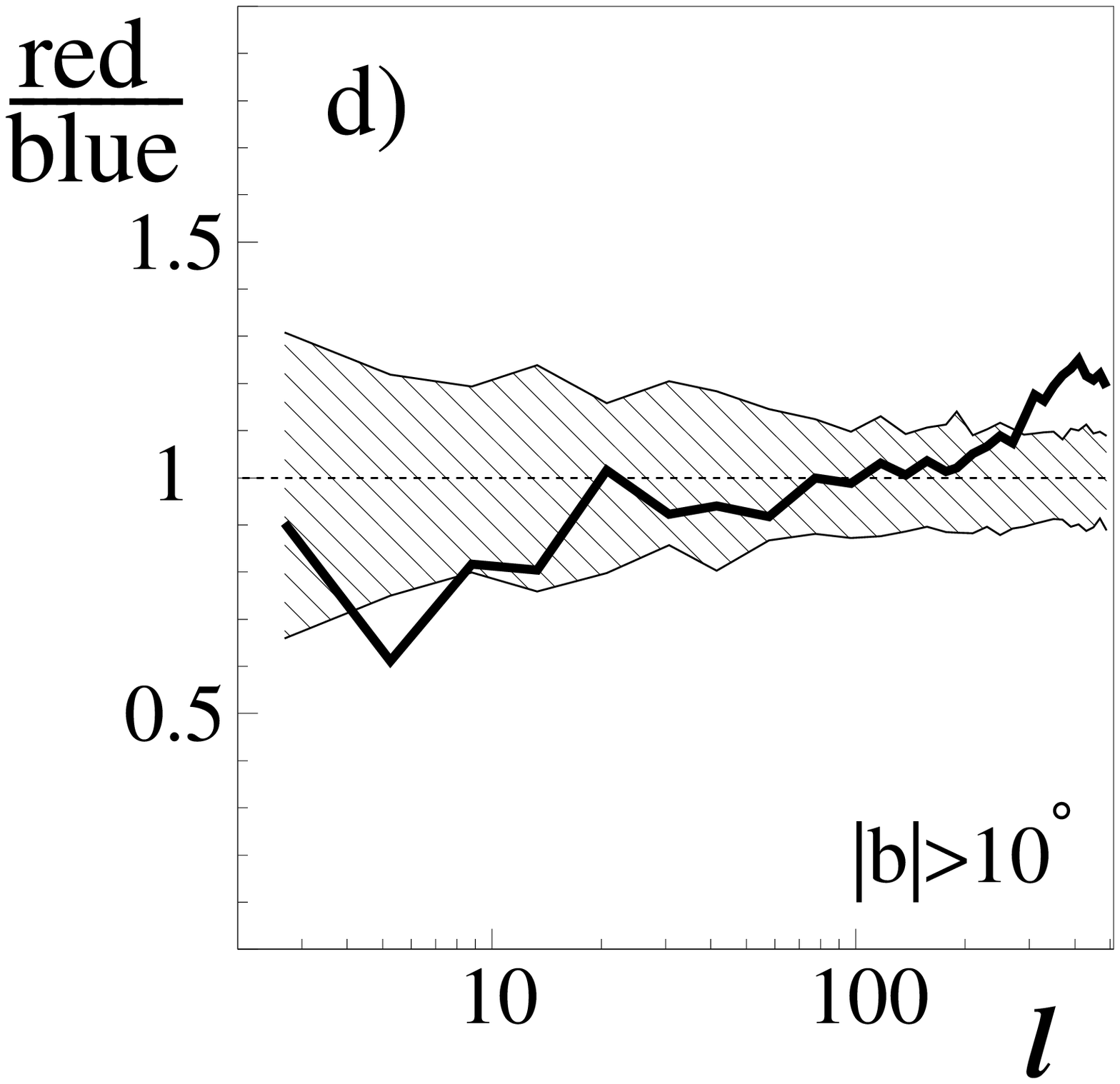}
}
\caption{Power spectrum for red (positive) and blue (negative
regions) and their ratios.  Studies of artifical universes show
essentially no red/blue differences.
show the limits 
 \label{f5} }
\end{figure}

Before moving to the possible 'cosmic-ray-connection' it is
necessary to refer to a feature that we have discovered which may
have significance.  We define points on the WMAP \citep{3}
 having
positive values as 'red' and negative values as 'blue'.  Power
spectra have been derived for the whole sky for red maps and blue
maps with the result shown in Figure~\ref{f5}a. Figure~\ref{f5}b
shows the ratio. Figures~\ref{f5}c and \ref{f5}d show the results
for $|b|>10^\circ$.  It is evident that in both cases,
$|b|<10^\circ$ and $|b|>10^\circ$, there is a systematic
difference in the shape of the power spectrum between observation
and expectation.

Comparison with the 'limits' for the artificial universes shows
two main features:

\begin{itemize}
\item [i)] a clear excess of red over blue at the highest
$\ell$-values;
\item[ii)] a less significant deficit for
$\ell <
10$.
\end{itemize}
The former can immediately be
attributed to the contribution from discrete sources.  The
significance of the important deficit at low $\ell$ needs
discussion.

Insofar as the power levels at $\ell$ values exceeding
$\Delta \ell \sim 5$ are 
largely independent, there are two
values (at $\ell =5$ and 10) in this vicinity which are in the
region of the lower limit.  If this limit corresponds to the one
standard deviation level then the chance probability is
$\simeq$ 7\%. The significance for cosmology of the
difference at low $\ell$ will be considered later.

Of greater importance is the observation of the clear slow increase
in ratio with increasing values of $\ell$.
This increase is due to increased
'foreground' in terms of both Galactic and Extragalactic
('discrete' sources) components. Inspection of artificial 
Universe results
shows that the chance probability is less than about 3\%.
Taking low
$\ell$-values as a datum, and assuming that the total 
foreground affects only
the red points, the result is that for the important 
one-degree peak the
extra contribution is about 10\%. Thus, from this argument, 
the power of the
peak should be reduced by this amount.

\section{Cosmic Ray -- CMB correlations}\label{iii}

\subsection{General Remarks} \label{iiiA}

As remarked earlier, the presence of a cosec $|b|$ variation in
the mean CMB temperature at each and every WMAP temperature from
22.8 to 93.6 GHz, \cite{8} shows that there is a definite
contribution to the uncleaned maps from Galactic -- 'thin disk'--
mechanisms.  These are generally regarded as: electron synchrotron
radiation, free-free radiation and dust.
Insofar as the
first mentioned is due to cosmic ray electrons and the dust is
heated, in part at least, by CR protons, we expected 'cosmic rays'
in general to be implicated in the foreground problem.  It is true
that subtracting off a cosec $|b|$ dependence, and arranging that
the mean CMB temperature is 2.74K, removes much of the CR-related
signal but some inevitably remains. Presumably, the residuals
remain on a variety of angular scales and a variety of magnitudes.
Insofar as our knowledge of the distribution of cosmic ray
intensity and gas density (of all densities, including that in
dust clouds) -- in 3-dimensions-- is very limited, a determination
of the likely CR-related contribution is very difficult.  Some
general points can be made, however, as follows:

\begin{itemize}
\item[i)] Lines of sight where the CR-intensity (of relevant
energy) is low might be expected to have low foreground
$\langle$CMB$\rangle$
(denoted $\langle$CMB$\rangle_f$).

\item [ii)] Lines of sight where the column density of gas is low
('Galactic chimneys' and other regions) and thus low CR
interaction should have low $\langle$CMB$\rangle_f$.

\item[iii)] Lines of sight crossing regions where the CR
intensity is high (SNR shells) \citep{21,22,33} would be expected to
have high $\langle$CMB$\rangle_f$.
\end{itemize}

Each of these regions will be considered in turn, as will an
all-sky correlation using the low energy gamma ray data from EGRET
\citep{3} and a more energetic component for $|b|<10^\circ$.

\subsection {Correlations in Steep CR-Spectra
Regions} \label{iiiB}

From the maps of electron and proton \citep{19,20} spectral
indices we have chosen regions covering about 10\% of the
sky area.  The results for a smoothing of 15$^\circ$ FWHM were
taken for both the proton and electron components in that we
expect CR 'variations' to be on this order of scale, corresponding
to typical SNR-induced large scale turbulence.  We expect the
contours for protons and electrons to be similar, but not quite
the same (differences of energies for the initiating particles,
electrons and protons, will cause differences in diffusion
coefficients). An estimate of the difference can be made, as
follows. The mean electron energy is about 1~GeV for the electrons
responsible for the detected gamma rays and about 10 GeV for the
protons so that the proton and electron escaping from their accelerating
SNR will be separated by of order 100 pc \citep{21} after a time of
10$^6$y (ten times the lifetime of the SNR).  At a typical
distance of 1~kpc the angular separation will be of
order 6$^\circ$. The somewhat disparate patterns of Figure~\ref{f5}
are thus understandable.

\begin{figure}
\centerline{
\includegraphics[width=15cm]{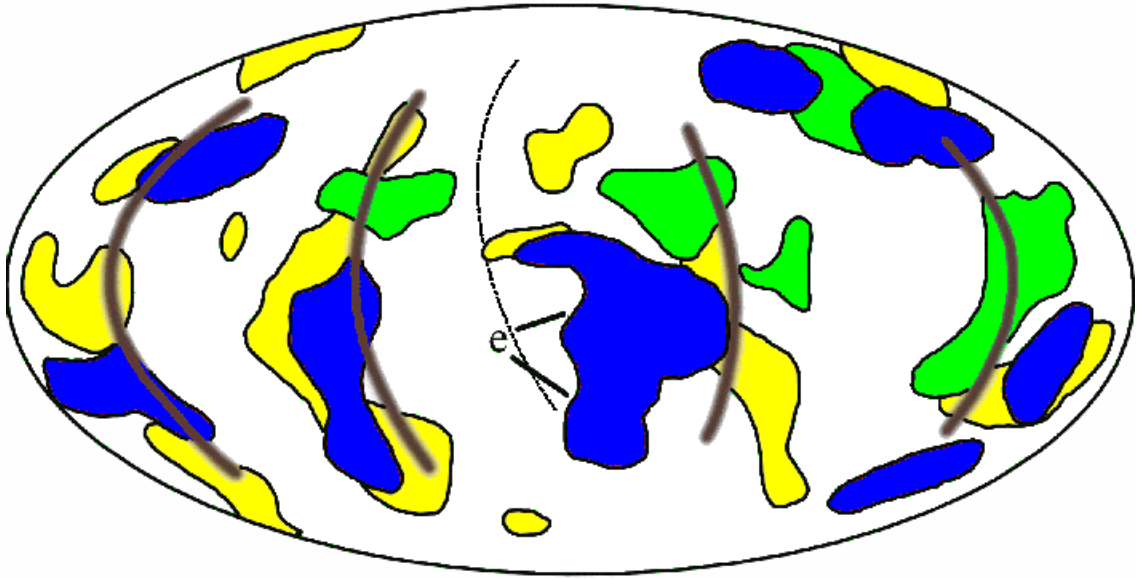}
}
\caption{Map of the sky (Galactic Centre at the centre, longitude
increasing to the left) showing regions of steep spectra for
protons 
and electrons \citep{19} 
and minima in the WMAP \citep{3}.
The thick lines represent low column densities of atomic hydrogen
(chimneys) and interarm regions. 'e' represents
 two regions of
steep electron spectra below the low CMB region. The ridge ('North
Polar spur') for the Loop I SNR is also shown 
as a dashed line.
 \label{f6}}
\end{figure}

In order to choose the most reliable regions of steep spectra -
and because we are not sure which of electrons and protons are responsible
- we have identified regions of common steep spectra for both
protons  and electrons in Figure~\ref{f6} (see also \cite{21}); these
should correspond to the most likely regions of maximum effect for
the CMB.  By 'maximum effect', here, we mean lowest CMB
temperature because steep spectra indicate low particle
intensities at the higher energies where we expect the extra
pseudo-CMB contribution to arise.

 Table~\ref{t1} indicates the
regions so identified and the corresponding values of
$\langle$CMB$\rangle$ 
It will be noted that all are negative. The significance
of the deficits will be examined shortly.

\subsection{Correlations for Galactic Chimneys} \label{iiiC}

Returning to the presence of chimneys of low column density of HI,
in which we have claimed steep spectra (and where we used limited
latitude regions above), an analysis has been made of the mean CMB
temperature along four major directions -- lines of constant
longitude (of width $\pm 6^\circ$) over their whole latitude range;
these directions do not always coincide with those of steep
spectra.

The mean temperature has been determined for
$10^\circ<|b|<60^\circ$, this being the nominal region over
which the chimneys can be identified.  The results are also given
in Table~\ref{t1}.

It can be seen that our expected reduction in mean CMB temperature
is observed in 6 or 7 out of the 8 regions.

These results are potentially important and they will be examined
in more detail.  In \cite{22}, where evidence was given for
correlations of steep spectra for low energy gamma rays with the
Galactic chimneys and low HI regions, a comparison was made
between $b_+$ ($+30^\circ$) and $b_-$ ($-30^\circ$).  It was found
that the $b_+$ results were better, those for negative latitudes
being inferior, particularly for $l =57^\circ$ and
$130^\circ$.  These results related to electrons.  With higher
energy gamma rays, however, where protons are the progenitors, at
negative latitudes, $l = 57^\circ$ was poor, but 130$^\circ$
was good.  The conclusion is that the anomalous positive result
for $b_-$ at $l = 57^\circ$ can, perhaps, be understood, not
least because inspection of Galaxy-wide HI column density maps
shows that the 'chimney' at negative $b$ and $l\sim
57^\circ$ is barely visible.

\subsection{Correlations for Other Low HI-Column Density
Regions}
\label{iiiD}

Another approach, 
 resembling the previous one, 
 is to look at
regions where the HI column density is low, irrespective of the CR
intensity and away from the HI-chimney.  The reason is that it may
be that it is the HI density itself that is important rather than
the CR intensity as such. Again, the results are shown in
Table~\ref{t1}.

\subsection{Correlations in High CR Intensity
Regions}
\label{iiiE}

If, as seems possible, it is CR interacting with gas that gives
rise to a pseudo-CMB contribution, then there should be an excess
of $\Delta T$ (CMB) in regions where the CR intensity is high.
Probably the best evidence for high CR intensities comes from
studies of gamma rays (typically above 100 MeV) from SNR shocks
\citep{31,22}.  It has been known for some years that the famous Loop
I SNR has somewhat higher gamma ray emission than expected \citep{31,22}.
Most came from the
well-developed ridge which is seen in radio, eg at 408 MHz, rising
along $b\sim$30$^\circ$ from the Galactic Plane.  The gamma ray
work just referred to also shows evidence for CR acceleration in
the shell of another SNR, Loop III, \citep{23} albeit at a lower
level.  Yet another SRN, Loop II, seems to be present, but rather
weak and is disregarded.

\begin{figure}
\centerline{
\includegraphics[width=15cm]{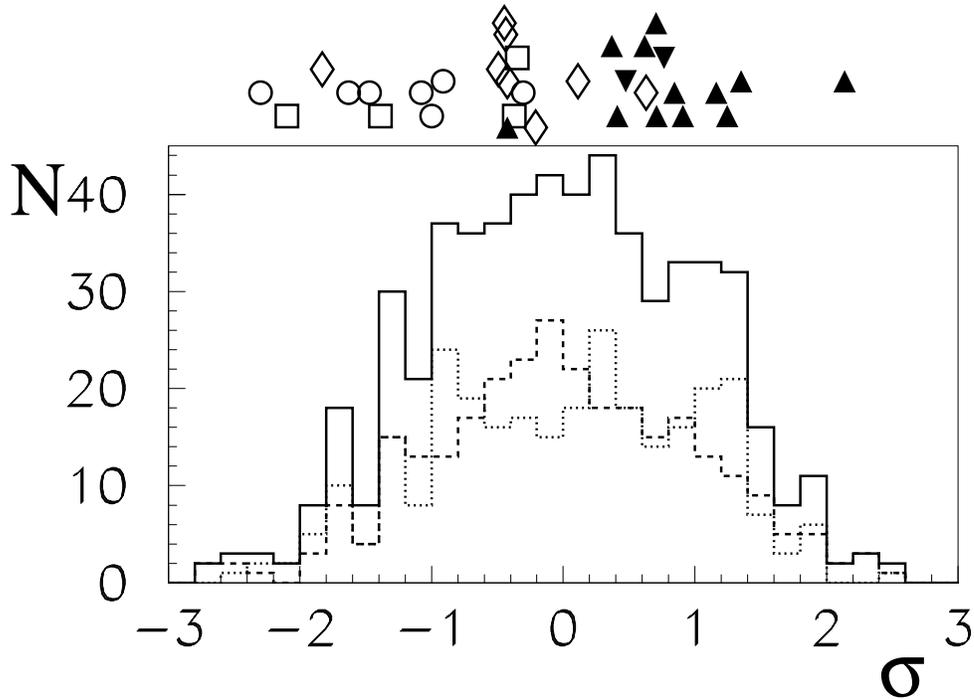}
}
\caption{\label{f7} Excesses and Deficits.  Values from Table~\ref{t1}
for the C-R associated excesses and deficits in comparison with
the frequency distribution of mean temperatures for similar
spatial regions shown by solid histogram 
(dashed and dotted histograms are for Noth and South hemispheres,
respectively).}
\end{figure}

Table~\ref{t1} shows the results for the leading edge of the Loop I
SNR (The North Polar spur), which is best identified in radio.
Also shown are values pertaining to the regions at
$|b|<10^\circ$ where the two Loops cross the Galactic Plane
(Figure~\ref{f7} shows the relevant CMB temperatures).  This region
is chosen so as to maximize any potential signal (positive or
negative) in view of the higher gas density here.

It is reassuring to note that the mean values of $\Delta T$ are
almost always positive.

It remains in this section to consider the statistical
significance of the correlation results so far.

Figure~\ref{f7} shows a histogram for the $\Delta T$ values all
over the sky.  The abscissa is in standard deviations and relates
to areas of the magnitude used in the analysis.  The symbols above
the plot are for the individual areas referred to above.  It is
evident that the divergences are, taken at their face value, very
significant.  Formally, the chance of getting such a situation is
$\sim10^{-4}$ for the negative excursions and $\sim 10^{-3}$ for
those in the positive direction.

It is, of course, possible to question the significance on a
number of levels: why choose the particular bin widths of
latitude, and the particular size of the low N (HI)
regions?  The answer is that they were chosen before the
results were known.  Nevertheless, it is best to conclude at this
stage that the correlations are 'very suggestive'.

\subsection{Whole-sky correlations of gamma rays and CMB}\label{iiiF}

The EGRET data \citep{4,3}
 have been used to search for correlations
over the whole Galaxy, although it is appreciated that the
statistical accuracy of the gamma ray data is not great.  Because
there is a strong latitude dependence of the gamma ray intensity
we study the CR-CMB correlation in narrow (but increasing with
$|b|$) latitude bins.  We concentrate on the highest energy (1
GeV) for which data of reasonable accuracy are available for the
whole Galaxy.

\begin{figure}
\centerline{
\includegraphics[width=15cm]{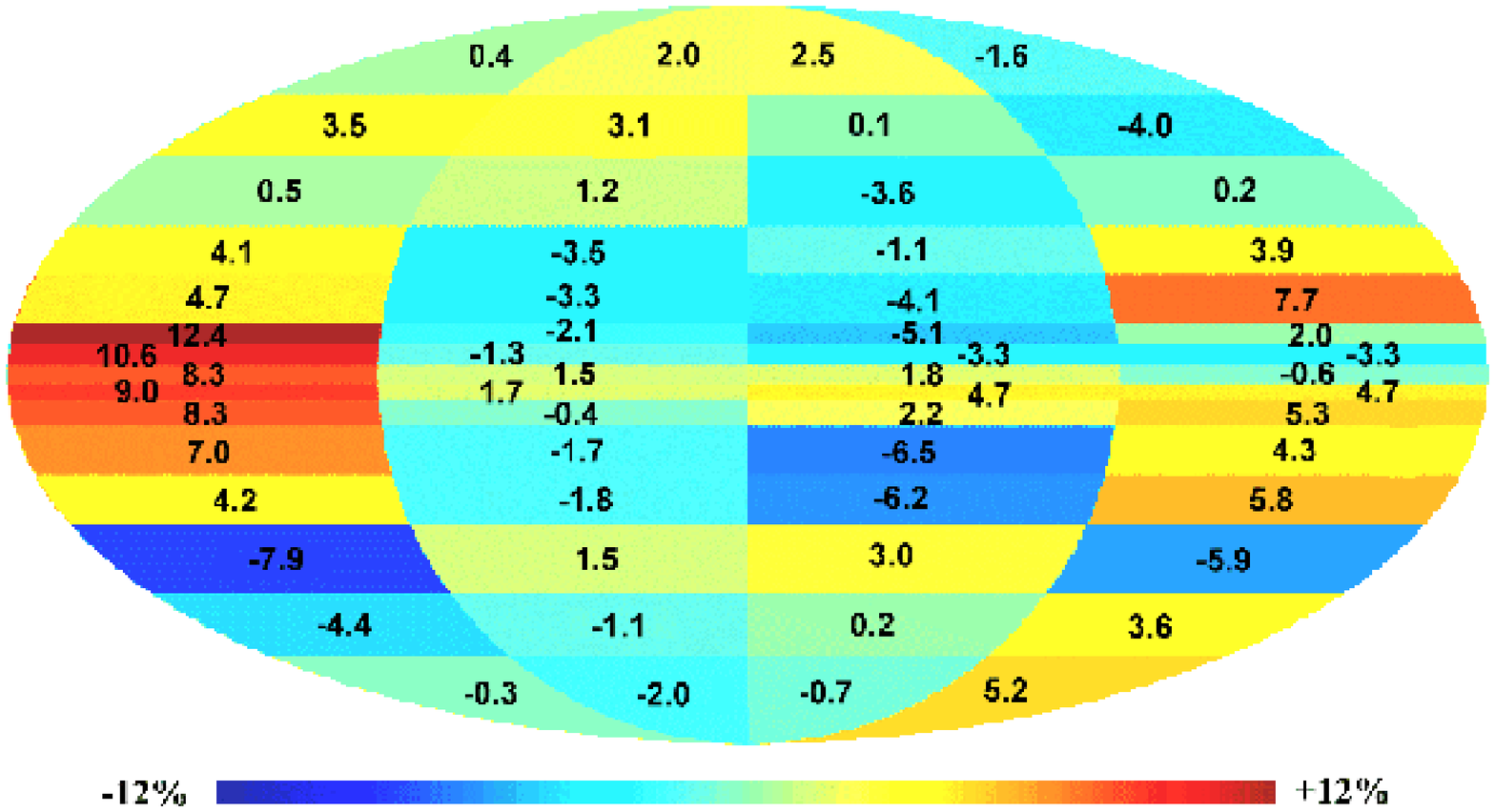}
}
\caption{Gamma rays above 1~GeV (\citep{4} and see \citep{2})
vs CMB temperature
\citep{3}.
Map of the correlation coefficients with the separate
Quadrants identified (Galactic scale as in Fig.~\ref{f6}). 
The units of the correlation coefficients
are 10$^{-2}$.  The latitude ranges are: $|b|<2^\circ$,
$2^\circ-6^\circ$, $6^\circ-10^\circ$, $10^\circ-20^\circ$,
$20^\circ-30^\circ$, $30^\circ-45^\circ$, $45^\circ-60^\circ$ and
above $60^\circ$, with similar divisions (below $b= -2^\circ$) for
negative latitudes. Although the statistical precision of any
individual coefficient is poor there are some important patterns,
most notably that there is a stronger correlation in the Outer
Galaxy, particularly in Quadrant 2 (in which all astronomical
indicators are stronger), and that there is a concentration in the
Galactic Plane. We think that the large gamma ray contribution
from Inverse Compton interactions and other factors, may 
have spoiled 
 the
correlation in the Inner Galaxy.
\label{f8}}
\end{figure}

The analysis has been made in two ways.  Firstly, by taking the
data in large, Quadrant size, bins in longitude and successive
latitude bins and
 by taking 10 degree radius regions round
successive pixels in the sky.  Both methods have their merits, and
demerits. Starting the with former, 
Figure~\ref{f8} shows a map of
the correlations in a manner such that the latitude -- and
Quadrant -- dependence can be sought.  It is evident that there
are both.  The Inner Galaxy has no overall positive correlation,
formally it is negative, but the Outer Galaxy has a positive
correlation, particularly in Quadrant 2. There is also a
pronounced dependence on Galactic latitude.  This behaviour is
surely suggestive of a relationship between the gamma ray
intensity and the average CMB temperature, although the
interdependence of bins is a worry.  It might be thought that one
could take wider intervals of latitude, near the Galactic Places,
but here the CR intensity is such a strong function of latitude
that artifacts can easily occur.

Before continuing the (largely negative) correlations in the inner
Galaxy will be considered further.  It is possible, in principle,
that the CMB radiation is absorbed in a particular geometrical way
by high temperature ionized gas in the Galactic Halo, but the
density and temperature needed are excessive.  Another source of
CR-intensity might be the answer: 
 Inverse Compton interactions
of electrons on starlight. 
 Another
possibility might be associated with the fact (Sect.~\ref{iiiC})
that there is the reverse gradient of the CMB power at latitudes
above 10$^\circ$.

We found Quadrant-dependent effects in Sect.~\ref{iiB} and it is
necessary to study Figure~\ref{f8} in some detail.  Although the
latitude regions are of unequal width they can be considered with
equal weight because, due to the strong $b$-dependence of the
gamma ray intensity, there are similar numbers of detected gamma
rays in each band.  In terms of strength of the correlation
coefficient, the net coefficients are:

Quadrant $1 = 4.8$; Quadrant 2 $+60.4$; Quadrant 3 $+27.3$; Quadrant 4
$-16.1$;

It is appreciated that adjacent latitude bins are not independent
but this dependence is not too large, in view of the observed
quite wide spread in coefficients from one bin to another.  Thus,
we conclude that there is some evidence for a correlation in the
Outer Galaxy.

For the range above $|b| = 20^\circ$, the summed coefficients
are:

Quadrant 1 $+ 4.7$; Quadrant 2 $-4.0$; Quadrant 3 $+4.3$; Quadrant 4
$-4.7$.

The differences here are evidently of no significance.

The positive correlation,(as evinced by this technique -- but see
later), is confined to $|b|<20^\circ$ and is predominantly in
the Outer Galaxy.  To what extent this result militates against a
CR-CMB correlation at $|b|> 20^\circ$ is not clear, in view of
the poor gamma ray statistics at these latitudes and the fact,
mentioned already, that rather than the average CMB temperature
that is important it is the temperature of the 'background' over
large angular scales (ie omitting the hot 'blobs') that should be
important.  This topic is taken up again later.

Returning to the whole sky, the North-South difference is also of
interest.  The net coefficient from the North (positive $b$) is $+28.0$
and that from the South is $+38.7$; again an S-excess.
Unfortunately, it is not clear how to work out the statistical
significance of this result, except to say that it is in the
correct sense. What can be done, however, is to rotate the CMB map
and rework the correlations.

This has been done and we have gone further to the extent of
examining the correlation between the CR-CMB correlation
coefficient and 
 $\Delta T$,
  this has been done because
we expect the coefficient to be higher in the regions of positive
excess temperature if, as is expected, CR (-like) effects give
additional signals.

The latitude region can be divided into two ranges, as usual:
$|b|<10^\circ$ and $|b|>10^\circ$.  For the former, all
3-latitude bands show significant positive signals (averaged over
$\Delta T$), for both N and S.  Furthermore, the profile of the
correlation of the coefficient, $C(\Delta T$), with $\Delta T$ is
similar in the N and the S.  In the rotated maps the similarity
between N and S is usually only strong in the region
$|b|<2^\circ$, and occurs because of the correlation between
adjacent bands -- it will clearly be a maximum for such a narrow
band, of width only 2$^\circ$.  Taking the average amplitude
alone, the chance of seeing positive signals of the magnitude seen
or  greater is estimated to be $<1$\%.  The shape similarity
(or coefficient vs $\Delta T$ in the N and the S) will reduce this
probability somewhat further, as will be demonstrated.

Continuing to higher latitudes, $|b| = 20^\circ - 30^\circ$
and $|b| > 60^\circ$ show small positive average coefficients
whereas $|b| = 30^\circ$ to 45$^\circ$ and $|b| = 45^\circ -
60^\circ$ show small negative correlations.  Before writing off
evidence for a genuine CR-CMB correltion coefficient vs
temperature at $|b|>20^\circ$,from this particular analysis,
however, we draw attention to three features:

\begin{itemize}
\item[i)] In the S region (our preferred hemisphere for extra
contributions) there is a distinct increase in correlation
coefficient with increasing $\Delta T$ in each of the 4 $|b|$
ranges.  In the rotated maps this happens in only 20\% of the
individual cases and in less than 10\% for each in a set of 4.
\item[ii)] From past arguments, we expect there to be a
correlation between the CR intensity and the 'power per unit solid
angle' at small $\ell$.  The average temperature
has a bigger contribution from the high temperature, spatially
condensed regions, regions which we do not expect to be affected
-- except in their base level -- by CR (-like) effects.  Taking
all data at $|b|>20^\circ$ we have examined the correlation of the
CR-CMB correlation coefficient with the mean power for the bin
(Figure~\ref{f3}, $\ell$=10). The coefficient is positive at the
15\% level. \item[iii)] The behaviour above $|b|$=45$^\circ$ has
relevance to the situation for $\Delta T_{\rm CMB}$ referred to
earlier and to the work considered in \cite{22}, viz the presence
of an axis of symmetry; inspection of the correlation
coefficients (Figure~\ref{f8}) for $|b|=45^\circ$ shows
pseudo-Galactic N and S poles at: $l,b=80^\circ,\: +55^\circ ;~
220^\circ, -55^\circ$. These directions are those of the maxima in
the two hemispheres beyond $|b| = 45^\circ$. They are reassuringly
close to those for $\Delta T$.
  The uncertainty in angle is,
again, approximately $\pm 25^\circ$.
\end{itemize}

The second approach, referred to above, has been made, using
smaller spatial regions for the correlations.  This involves
comparing the appearance of the genuine WMAP and those for
artifical universes. Two features in Figure~\ref{f8} are
considered by us to provide the evidence for a bona fide
correlation of CR with CMB: the excess in the Galactic Plane and
the presence of symmetry about the plane.

For the genuine map we derive the integrated length of longitude
for which the CMB temperatures are above the 'red' minimum
($\Delta T >
 300 \mu K$).  This is $115^\circ$ for the WMAP
and, for the artificial universe, the mean is $33^\circ \pm
17^\circ$.  Thus, formally, the WMAP
effect is at 5 standard
deviations.  The significance is almost certainly less than that
for a Gaussian distribution, but is it probably greater than
corresponding to a 1\% probability of being spurious.  For
symmetry we count the number of symmetrical regions of length at
least $\Delta l=10^\circ$.  There are 7 symmetrical regions
compared with 2.0 $\pm$ 1.4 for the artificial universes,ie a 3.6
standard deviation effect.  The corresponding chance probability
is almost certainly less than 5\%.

Continuing to higher latitudes, 
 we find, at $|b|$=30$^\circ$ and
60$^\circ$, integrated red regions significant to the extent of
2.8 sigma and 2.6 sigma in comparison with the artificial universe
results.

Our conclusion from this large scale CR-CMB correlation vs
'temperature' study is that it exists.

\subsection{Restricted region studies at the highest
energies}
\label{iiiG}

\begin{figure}
\centerline{
\includegraphics[width=10cm]{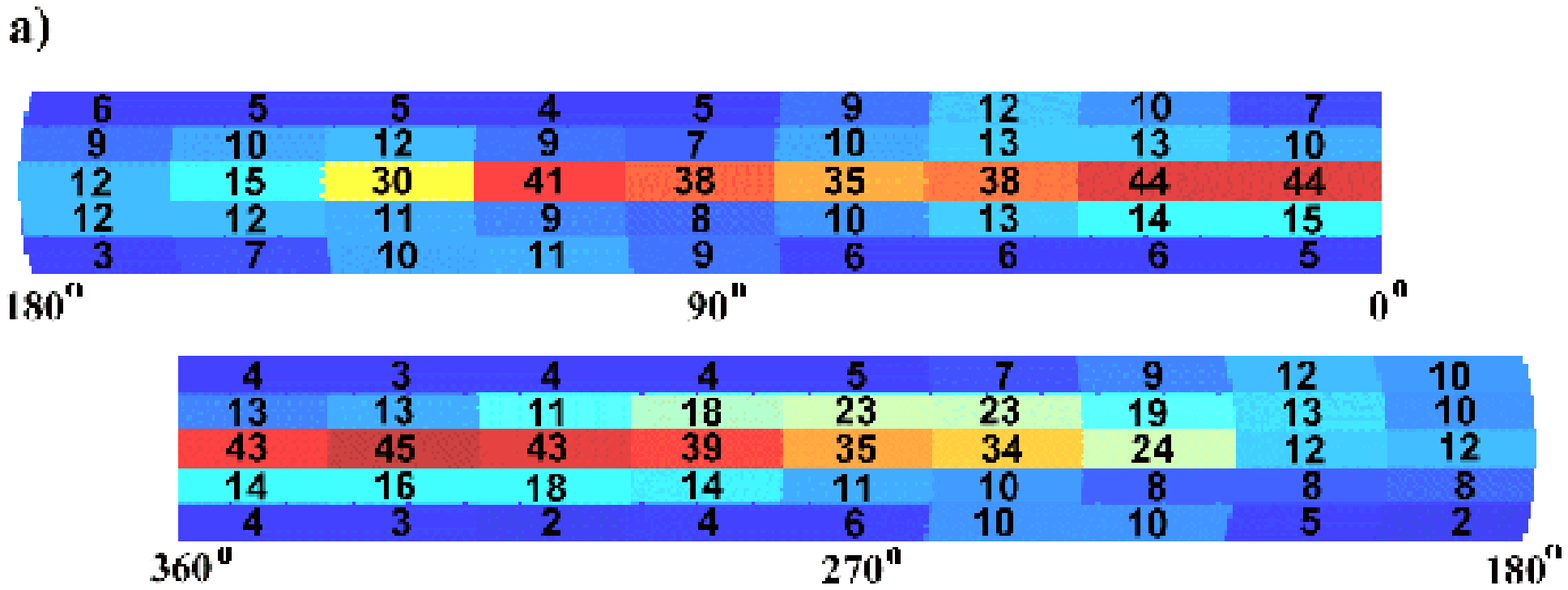}
}
\centerline{
\includegraphics[width=10cm]{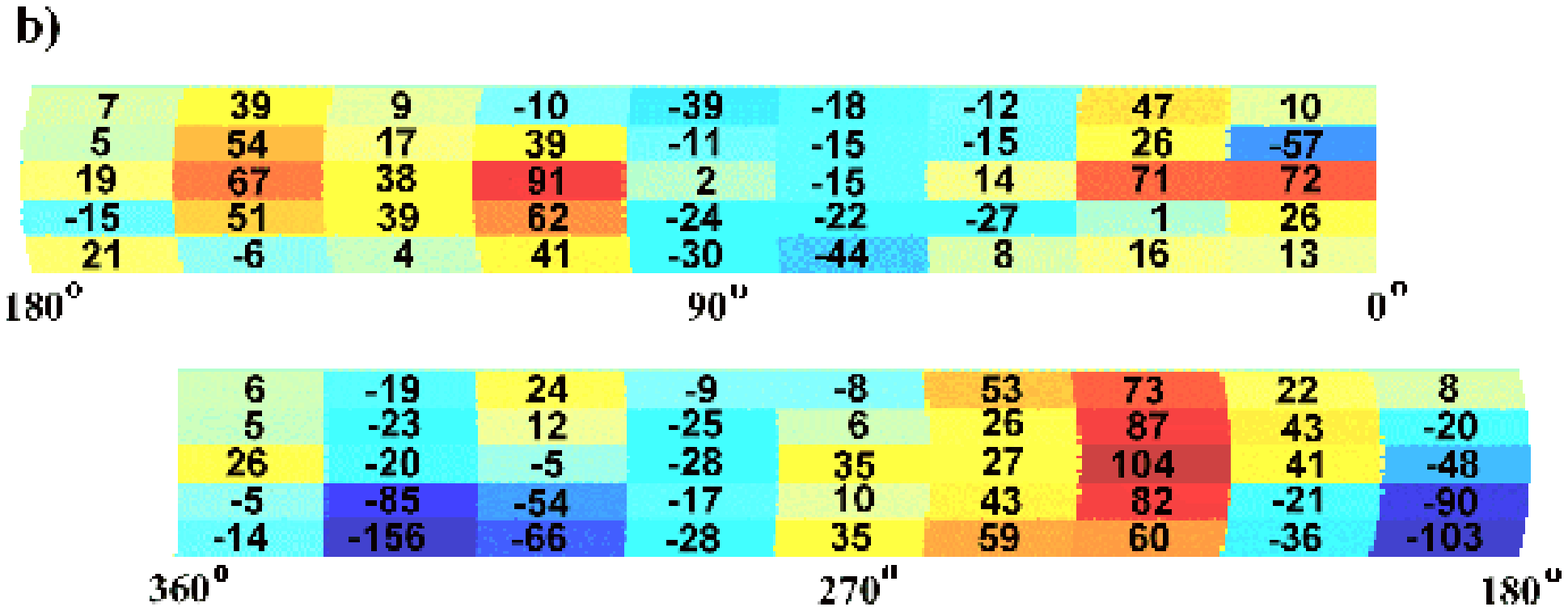}
}
\centerline{
\includegraphics[width=10cm]{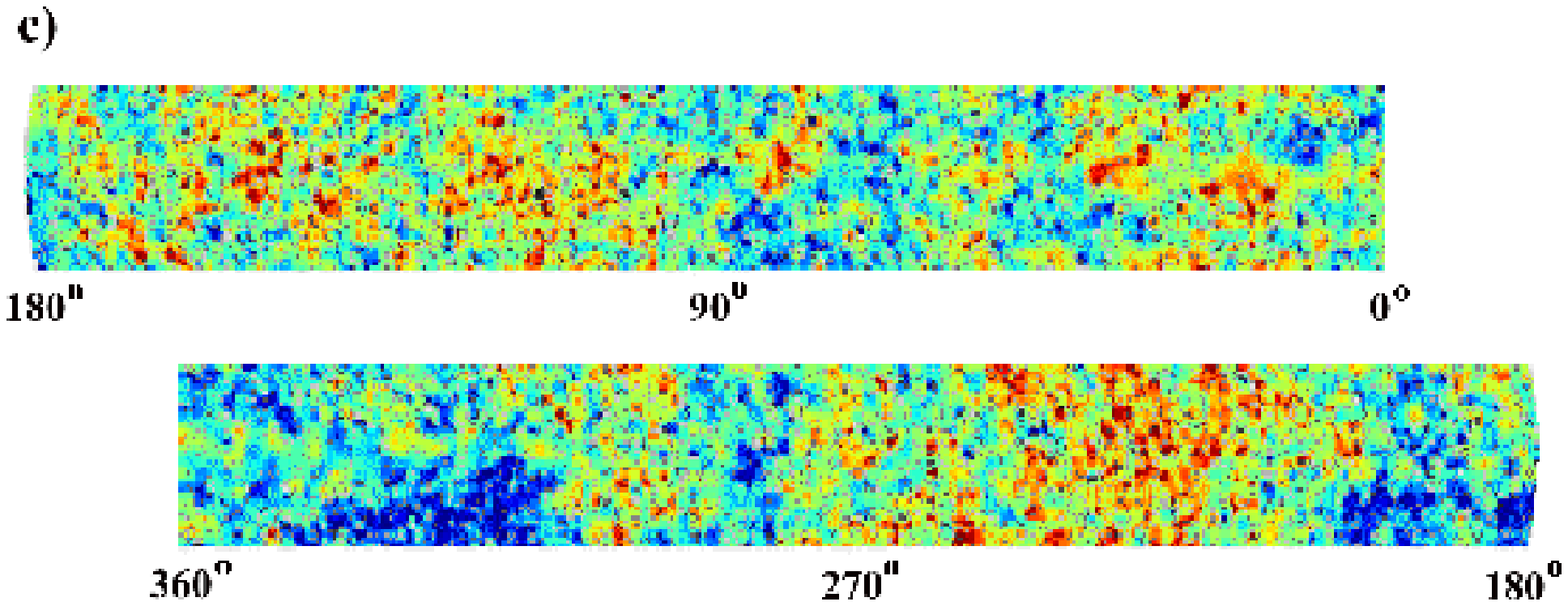}}

\centerline{
\includegraphics[width=6cm]{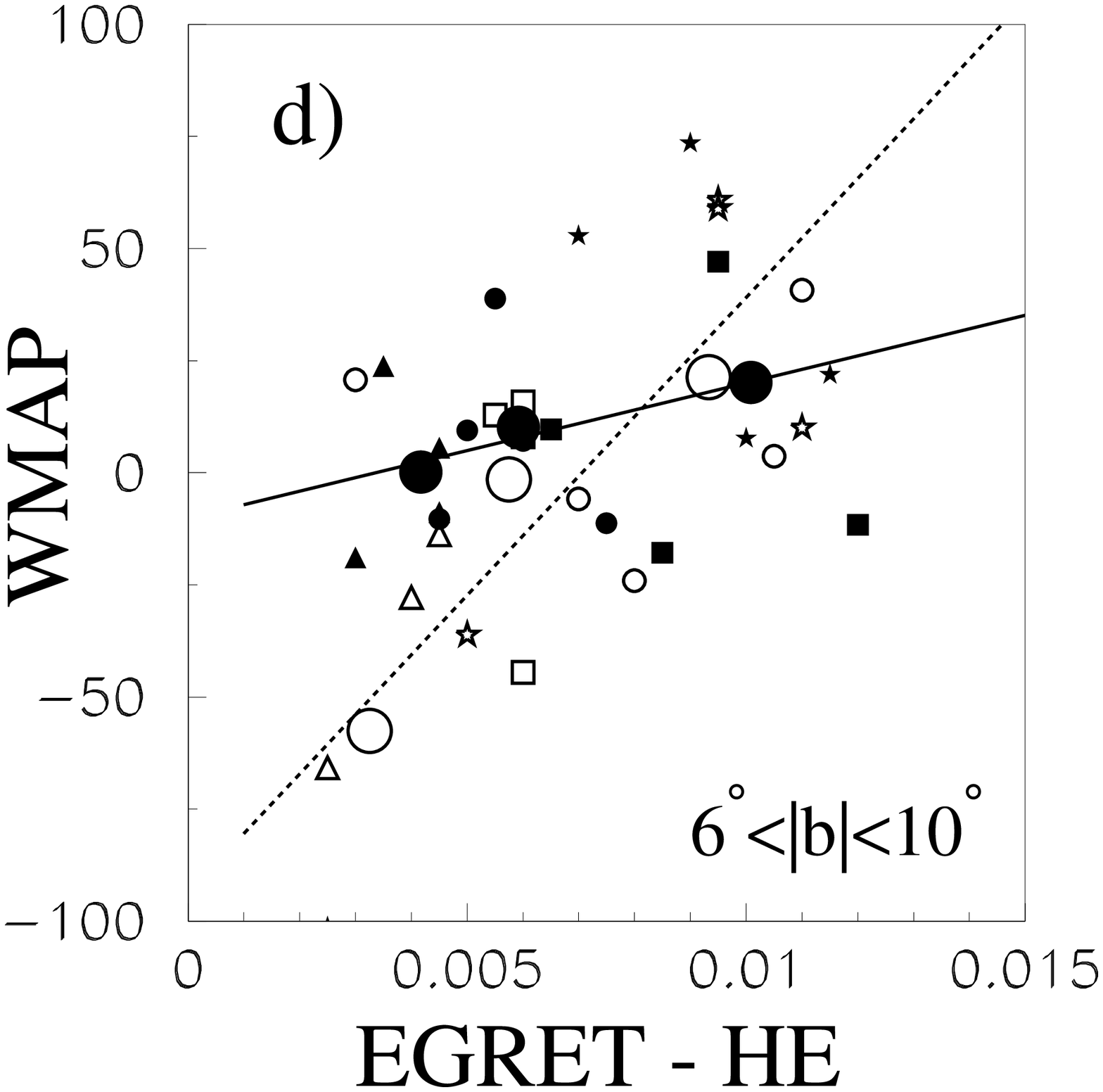}
\includegraphics[width=6cm]{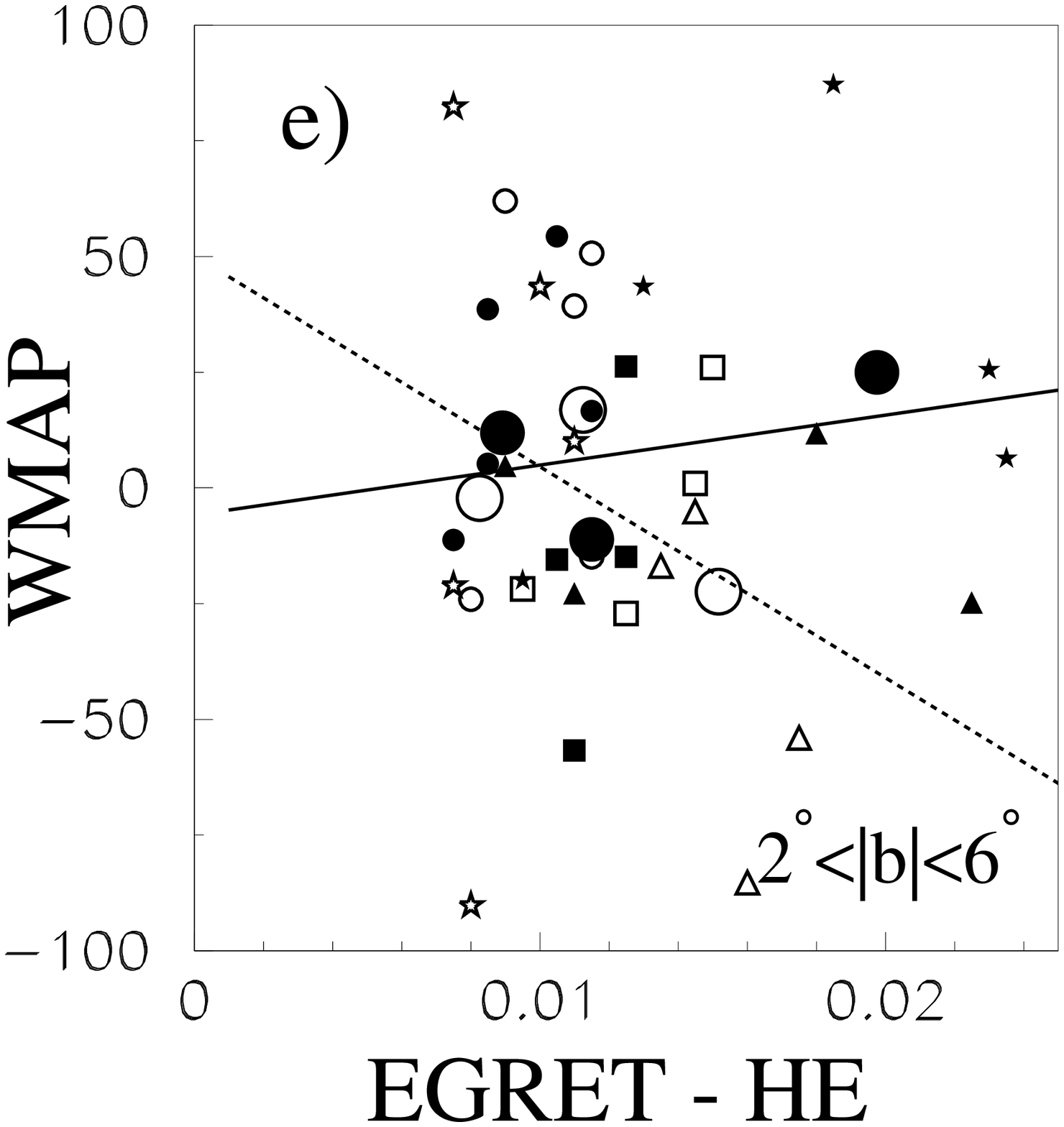}
\includegraphics[width=6cm]{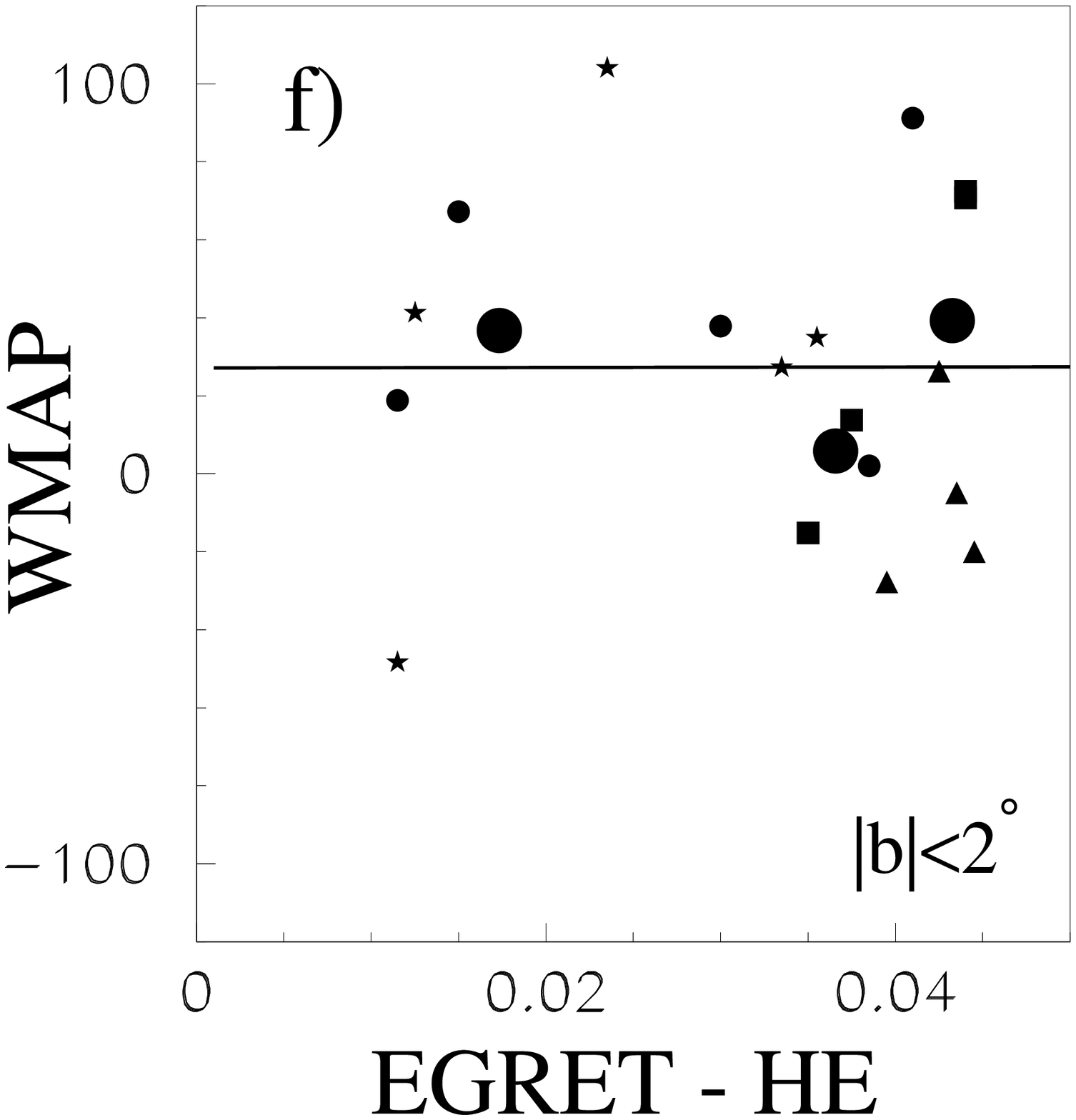}}

\caption{Gamma rays at 30~GeV
from \citep{4} 
 for the restricted latitude
range where adequate statistic
precision is available, viz $|b|<10^\circ$ 
vs
CMB mean temperature \citep{3}
The best-fit lines show that the correlation is rather strong for
the range $|b| = 6^\circ - 10^\circ$.
Open circles: Southern hemisphere, closed circles:
Northern hemisphere.
The large circles represent the data concentrated in 3 bins.
\label{f9}}
\end{figure}

EGRET data are available to 30 GeV for a restricted region,
$|b|<10^\circ$, and a correlation analysis has been made here.
Again, because of the strong $b$-dependence of the gamma ray
intensity, we take narrow bands of latitude. The results are shown
in Figure~\ref{f9}. Although the spread is large there is a clear
correlation of the high energy EGRET flux and the CMB excursions,
for the region away from the very lowest Galactic latitudes, viz
$|b| = 60^\circ - 10^\circ$.

We would not have expected correlations for $|b|<2^\circ$ (and to
a lesser extent for $2^\circ - 6^\circ$) where the lines of sight
are very long and where there is considerable 'confusion'. For
$6^\circ<|b|<10^\circ$, where we expect the best correlation
following the arguments contributing to Figure~\ref{f7}, the slope
is 2.2 standard deviations from 
 zero taking all the data and
significantly more for the (CR-favoured) S hemisphere, with
smaller significance for the North.

\subsection{Indirect studies of a CR-CMB Correlation for
$|b|>10^\circ$}
\label{iiiH}

Circumstantial evidence comes from N-S, $b$-dependence effect and
Quadrant differences.

For CR-related effects we would expect:

A southern excess; mean for Q2 $>$ mean for Q3; a fall-off with
increasing latitude, where the quantity under examination is
either the power at low $\ell$ or the mean temperature.

The results are, for Q1, 2 and 3 (Q4 is omitted, because of the
'big blue hole')

\begin{itemize}
\item[1.] Power at $\ell$=10. S/N$>$1, Q2$<$Q3, Figure~\ref{f2}; fall-off
with increasing $|b|$ (figure~\ref{f3})
\item[2.] Mean CMB
temperature.  Southern excess S-N $= 30 \mu K$, similar for both Q2
and Q3. Q2 $-$ Q3 $= +60 \mu K$. A fall-off of $\langle T\rangle$ with increasing
$|b|$ from as far as $|b|=45^\circ$ (from right down to $|b| =
2^\circ - 6^\circ$).
\end{itemize}

Evidently, most of the features expected do occur, the only
exceptions being the reversal of Q2$>$Q3 seen for the $\ell$=10
power intensity and the termination of the fall in $\langle T \rangle$ with
increasing $|b|$ at 45$^\circ$.  The former is considered later in
Sect.~\ref{iv}. The last mentioned is interesting in its own right. It is seen
in all Quadrants (Figure~\ref{f4}) and usually for both N and S. It
is germane to remark that the 408 MHz and 5 individual WMAP
frequency plots all show an upturn (or constancy) of intensity at
$|b|$ = 60$^\circ$, in order of increasing $|b|$.  In our case,
for the N, the increase (for Q's 1,2 and 3) is 51 $\mu K$ whereas
for the South it is $+3 \mu K$; in the 408 MHz and WMAP
frequencies, 4 out of the 5 sets of data which show an increase
have a bigger effect in the North. In the power spectrum,
$\ell$=10, data (Figure~\ref{f4}) there is an increase in 4 out of
6 cases although that for the South exceeds that from the North.

\subsection {Summary of the CR-related - CMB
correlations}\label{iiiI}

It is concluded that for $|b| <10^\circ$ there is good evidence
for a correlation of some form; each and every analysis shows an
effect.  The overall chance probability is
 less than about
2\%.

In the range $|b| = 10^\circ$ to $20^\circ$ there is ambiguity,
epitomized by the systematically low intensities in the power
spectrum for both $\ell$ = 10 and $\ell$=100.  This is reflected
in $\langle \Delta T \rangle$
 only if Q4 is included.  Without it,
the mean temperature for $|b| = 10^\circ - 20^\circ$ is only a
little above zero.

Moving to $|b|>20^\circ$ there is evidence for a correlation of CR
with either power intensity (at $\ell=10$) or $\langle\Delta T
\rangle$
 but the most direct evidence comes from the deficit in
$\langle\Delta T \rangle$ 
 in the special regions (Galactic
chimneys and regions of steep CR spectrum) and the excess in the
SNR shock regions. Insofar as the special regions cover the range
$|b|= 10^\circ - 20^\circ$, too, there is evidence for
correlations here, too, despite the ambiguities referred to above.
The overall chance probability is, conservatively, at the 2\%
level.

\section{Combination of External and Internal studies of the
WMAP}
\label{iv}

By `external' we mean the comparison of the WMAP data with cosmic ray parameters
and by `internal' we mean a comparison of WMAP data within itself, eg N-S asymmetry,
$b$-dependence.

The largest effect concerning non-uniformity relates to low
latitudes, typically $|b|<10^\circ$, or somewhat wider.  Both
internally and externally there are differences at low latitudes;
a Galactic origin is surely very likely here which may, or may
not, be CR-related.   The main question to be addressed is ``are
there foreground effects beyond $|b|=10^\circ$ or, in view of the
anomalous region $|b| = 10^\circ - 20^\circ$, beyond 20$^\circ$

This aspect will now be addressed in some detail.

The evidence favouring CR-related effects at $|b|>20^\circ$
 can be
summarised as follows:
\begin{itemize}
\item[i)] The deficit in the mean WMAP temperature associated
with the regions of steep CR spectra and low HI column densities
are quite significant, and extend to $|b|\approx 60^\circ$.
\item[ii)] The excess for the North Polar Spur similarly extends
to $b=+60^\circ$.  It, too, is significant.
\item[iii)] The large
scale CR-CMB correlation adds a little weight - at least at the 3\%
significance level (although the absence of much
correlation in the Inner Galaxy is a worry).
\item[iv)] The N-S
symmetry is large at small $\ell$ (for the power spectrum) and
continues to $|b|\sim 55^\circ - 60^\circ$, after which it falls.
This is mirrored by the 408 MHz signal and by the individual WMAP
data for the various frequencies.  It is generally agreed that the
$|b|$-dependence of the 408 MHz and the individual WMAP data sets
is due to CR-related Galactic foreground and the low $\ell$ power
spectrum asymmetry presumably has the same cause.
\item[v)] The
form of the actual latitude dependence of various parameters
presumably also has the same cause.  For the five frequencies the ratio
of $\Delta T(20^\circ-30^\circ)$ to $\Delta T(90^\circ)$ is
2.1 (it falls slowly with increasing frequency from 2.3 to 1.9).
Our ratio for the power at $\ell =10$ is 1.7, ie 1.3 for the root
mean square temperature.  The ratio for larger $\ell$-values will
be smaller.  The relationship between the longitude-averaged
$\Delta T$ values for the five frequencies and that for
$\sqrt{\Delta T^2}$ for small $\ell$ ($\ell=10$) is not completely
clear but if the $\Delta T$ fluctuations are a constant fraction of
the total DC signal then the '2.1' is to be compared with the
'1.3'.  Taken at its face value it would mean that the maps were
cleaned to the extent of a factor 4 - and a greater factor at
higher $\ell$ (eg by $\sim8$ at $\ell=30$, using Figure~\ref{f1}d).
These factors relate to $|b|>20^\circ$ where most of the CMB
signal used for cosmological analyses resides.  A lack of cleaning
to the extent derived (25\% of the signal at $\ell=10$
being CR-related noise) at $\ell\sim 10$ is clearly
serious.
\item[vi)] At the highest latitude the consistency of the
maximum intensity, from one parameter to another, adds confidence.
\item[vii)]  A comparison of both $\langle T\rangle$ and power (for $\ell \sim 10$) for
Q2 and Q3 gives understandable results in terms of Galactic
phenomena.
\end{itemize}

\section{Application to Cosmology}\label{v}
\subsection{General Remarks}
\label{vA}

In the absence of clear evidence as to the exact origin of the
foreground responsible for the various anomalies in the CMB map it
is not possible to correct for it and make an updated cosmological
interpretation.  However, some progress can be made by looking in
detail at three important parameters:
\begin{itemize}
\item[-] the
height of the 'one-degree peak';
\item[-] the $\ell$-value of the
peak, and
\item[-] the amplitude of the power spectrum at low $\ell$
($\ell<10$).
\end{itemize}

\subsection{The height of the one-degree peak}\label{vB}

\begin{figure}
\centerline{
\includegraphics[width=9cm]{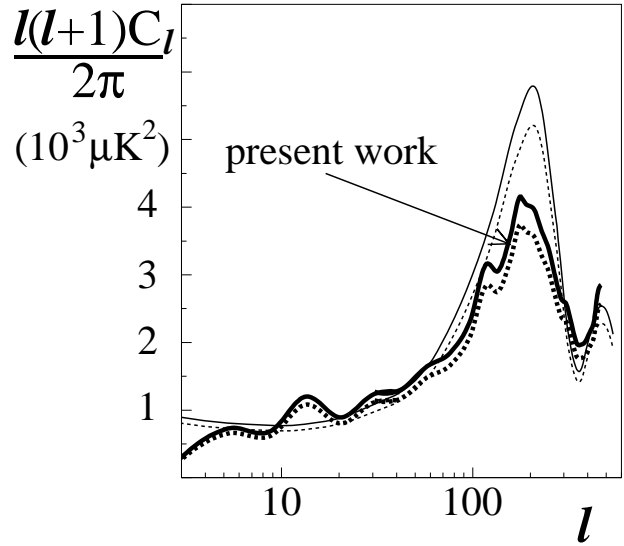}}
\caption{The upper curve is a standard result from WMAP
\citep{1,9},
the next one down is 10\% below this and the 
next pair are 
from our own analysis of the WMAP data for the whole sky presented
in \citep{4} (see \citep{2}).
\label{f10}}
\end{figure}

Figure~\ref{f10} shows the power spectrum for the whole data used
by us.  Also shown is what is generally regarded as the standard
form as given by the WMAP workers themselves (and a curve
10\% below this). We note that 'our' value is low.
Interestingly it is nearer to the conclusions drawn from pre-WMAP
observations (see \cite{17}). Specifically, the heights of
$\ell(\ell+1)C_\ell/2\pi(\mu K^2)$ are:
\begin{itemize}
\item[-] WMAP group \citep{30}   5700
\item[-] previous work
[\citep{17}]  $4925 \pm 175$
\item[-] present work  4250 (whole sky)
3490 ($|b|>10^\circ$, scaled up).
\end{itemize}

The difference in height of the peak between our value using the
data of \cite{3}
 and the datum \cite{30} is seen to be at least 25\%
and therefore serious.

Turning to the variations from Quadrant to Quadrant and N to S,
Figures~\ref{f1} and \ref{f2} indicate that these are (at $\ell
\sim200$) of order 10\%; similarly, Figure~\ref{f5} suggests
that the observed peak is some 10\% too high, in the sense
that the red (with extra signals) and blue (without) differ such
as to give this difference.

There are two ways forward at this juncture, the first is to
assume that our derived value of 3490 (less 10\%), ie 3140
$\mu K^2$ is correct and the other that there is a systematic
displacement in the scale of the result derived using the cleaned
data of \cite{3},
 and that we should take 90\% of the WMAP
groups' result \cite{30,18}, viz $0.9\times 5700 = 5130\mu K^2$.

The implication of a low peak value is most likely that
$\Omega_m$, the mass content of the universe, should be reduced,
although we appreciate that other variants to the cosmological
parameters can be taken.  Nevertheless, a reduced peak height is
most easily achieved by increasing $\Omega_m$ (or specifically
$\Omega_mh^2$) \citep{24}.  A peak height of 3590 $\mu K^2$ would
indicate $\Omega_mh^2=0.40$, a very high value, and 5130 $\mu K^2$
yields $\Omega_mh^2=0.20$, itself significantly higher than the
'standard value'.

\subsection{The $\ell$-value of the one-degree peak}\label{vC}
Less susceptible to systematic errors is the $\ell$-value of the
peak.  Conventionally it is at $\ell_A$ = 220 but it is apparent
that our result is systematically lower.

Values have been derived for $|b| > 10^\circ$ for the Quadrants,
by fitting a 'standard shape' of W($\ell$), derived from the whole
data.  The values are:
\begin{itemize}
\item[-]  Quadrant 1  $205\pm 10$;
\item[-]  Quadrant 2 $ 180 \pm 10$;
\item[-]  Quadrant 3 $190\pm 10$;
\item[-]  Quadrant 4 $210 \pm 10$
\end{itemize}
\noindent giving a best-estimate of $196 \pm 5$.  This value is
$24 \pm 5$ lower than the conventional 220.  Using the expression
$\Delta\ell_A/\ell_A$ = -0.24 $\Delta\Omega_mh^2$ \citep{24} and a
datum of $\Omega_mh^2$ = 0.15, we have $\Delta\Omega_mh^2$ = 0.17,
ie $\Omega_m h^2$=0.22.

An alternative approach is by way of Figure~\ref{f5}.  Again,
using the blue spectrum we have $\ell_A = 200\pm 10$, ie
$\Delta\ell_A = -20 \pm 10$ and $\Delta\Omega_m h^2 = 0.06$, ie
$\Omega_m h^2 = 0.21$.  The estimated error is $\pm 0.05$.

\subsection {The situation at low $\ell$-values}
 \label{vD}

It is well know that there is a loss of power below $\ell=10$ in
comparison with conventional expectation \citep{30} and this has
given rise to speculation of 'new physics' (eg \cite{26}).  One
possibility is that it is 'mere' cosmic variance that is
responsible and it is true that this is large at low
$\ell$-values.  The big differences from Quadrant to Quadrant,
both for low and high latitudes, have relevance here; they show
that there may be genuine differences.  However, these may be due
to CR-related foregrounds insofar as the deviations from one
Quadrant to another are bigger in the region $|b|<10^\circ$.

At first sight there is evidence for an even greater fall off of
power at low $\ell$ with increasing latitude so that, in the
absence of a cosmic-variance-inspired effect, the anomaly might be
thought to be secure.  However, we are struck by the results in
Figure \ref {f5} which shows a near-normal power law intensity at
low $\ell$-for the blue regions, ie the favoured set of data.
Although the peak of the blue curve is low, from $\ell=100$
downwards, the form of the blue curve mirrors that of the
conventional prediction all the way down to the lowest values of
$\ell$.  A possibility is that the form of the foreground
contribution is such as to reduce the apparent power at low
$\ell$-; we remember that the total temperature variations are
normalized to zero.

\section{Conclusions}

The major conclusion from this work is that there are features of
the WMAP of the fluctuations in the CMB that are not related to
the early universe, but are due to a 'foreground'.  Some 
of the foreground at least
is surely of Galactic origin and it appears to exist at all
Galactic latitudes.

Large scale asymmetries (also seen by others) exist on scales of
North vs South and Quadrant to Quadrant.  Our understanding of
them is limited but guidance seems to come from correlations of
the mean temperature, and intensity of the power spectrum, with
cosmic ray features.

The implications for cosmology are complex, but some remarks can
be made, as follows:
\begin{itemize}
\item [1.] Presently derived
cosmological constants are more uncertain than usually
appreciated.
\item [2.] If it is assumed that the only parameter to
be adjusted is the universal mass fraction $\Omega_m h^2$ then
there is evidence for it being larger than the canonical 0.15. Two
of our analyses suggest that it is $\sim$0.22 $\pm$0.05.
Another analysis gives a value as high as 0.40.  It is premature,
however, to put much weight on this very high value. Nevertheless,
a value higher than the canonical $\Omega_m h^2$ = 0.15 is
preferred.  Of course, variations of the values of other
cosmological parameters are possible, but such variations would
also be of great interest.
\item [3.] The situation at low $\ell$ is
interesting and our analysis suggests that the more conventional
view may prevail, viz that the apparent missing power may be an
artifact.
\end{itemize}

\section*{Acknowledgments}
 The authors are grateful to W.J. Frith for helpful comments and
to M. Tegmark for helping us to access the WMAP data.

Some of the results in this paper have been derived using the
HEALPix \citep{28}.



\begin{table}
\caption{Average CMB $\Delta T$ for special regions\label{t1}}
\begin{tabular}{cc|r|r}
\hline
\multicolumn{4}{c}{steep electron/proton spectra}\\
\multicolumn{1}{c}{$b$}&\multicolumn{1}{c|}{$l$}&\multicolumn{1}{c|}{$\langle \Delta T \rangle$ ($\mu K$)}&
                          \multicolumn{1}{c}{($\sigma$)}\\
  $(50 \div  55)$&$ (54\div  66)$&$       -53.9    $&$  -1.714$\\
  $(22 \div 27)$&$  (54\div  66)$&$       -11.1    $&$  -0.266$\\
  $(33 \div 38)$&$ (124\div 136)$&$       -46.6    $&$  -1.492$\\
  $(45 \div 50)$&$ (219\div 231)$&$       -39.5    $&$  -1.317$\\
 $(-25 \div-10)$&$ (304\div 316)$&$       -47.3    $&$  -0.903$\\
  $(10 \div 25)$&$ (304\div 316)$&$       -24.4    $&$  -0.958$\\
 $(-45 \div-15)$&$ (180\div 200)$&$       -31.5    $&$  -0.596$\\
\hline
\multicolumn{4}{c}{low HI column density}\\
\multicolumn{1}{c}{$b$}&\multicolumn{1}{c|}{$l$}&\multicolumn{1}{c|}{$\langle \Delta T \rangle$ ($\mu K$)}
&                          \multicolumn{1}{c}{($\sigma$)}\\
  $(40 \div 60)$&$  (90 \div  140$&$       -17.6   $&$   -0.391$\\
  $(40 \div 50)$&$  (60 \div 75)$&$       -16.1   $&$   -0.503$\\
  $(25 \div 60)$&$ (210 \div 270)$&$       -17.6   $&$   -0.361$\\
 $(-60 \div-45)$&$ (310 \div 15)$&$       -28.9   $&$   -0.131$\\
\hline
\multicolumn{4}{c}{Galactic chimneys}\\
\multicolumn{1}{c}{$b$}&\multicolumn{1}{c|}{$l$}&\multicolumn{1}{c|}{$\langle \Delta T \rangle$ ($\mu K$)}&
                          \multicolumn{1}{c}{($\sigma$)}\\
  $(10\div  60)$&$  (51 \div 63)$&$         2.1   $&$    0.110$\\
 $(-60 \div-10)$&$  (51\div  63)$&$        24.0   $&$    0.630$\\
  $(10 \div 60)$&$ (304 \div316)$&$        -7.9   $&$   -0.423$\\
 $(-60\div -10)$&$ (304 \div316)$&$       -18.8   $&$   -0.494$\\
  $(10 \div 60)$&$ (124 \div136)$&$       -34.4   $&$   -1.830$\\
 $(-60 \div-10)$&$ (124 \div136)$&$       -17.0   $&$   -0.447$\\
  $(10 \div 60)$&$ (219 \div231)$&$        -4.0   $&$   -0.211$\\
 $(-60 \div-10)$&$ (219 \div231)$&$       -16.7   $&$   -0.438$\\
\hline
\multicolumn{4}{c}{`Loops'}\\
\multicolumn{1}{c}{$b$}&\multicolumn{1}{c|}{$l$}&\multicolumn{1}{c|}{$\langle \Delta T \rangle$ ($\mu K$)}&
                          \multicolumn{1}{c}{($\sigma$)}\\
  $(74\div 86)$&$ (   9\div  21)$&$      21.4    $&$   0.618$\\
  $(64\div  76)$&$ (   9\div  21)$&$        42.0    $&$   1.350$\\
  $(54\div  66)$&$ (  24\div  36)$&$        26.1  $&$     0.705$\\
  $(44\div  56)$&$ (  25\div  37)$&$        -8.9  $&$    -0.423$\\
  $(34\div  46)$&$ (  26\div  38)$&$        13.3  $&$     0.411$\\
  $(24\div  36)$&$ (  26\div  38)$&$        30.4  $&$     0.945$\\
  $(14\div  26)$&$ (  24\div  36)$&$        61.5  $&$     2.138$\\
  $( 4\div  16)$&$ (  22\div  34)$&$        39.4  $&$     1.161$\\
  $(-6\div   6)$&$ (  18\div  30)$&$        32.6  $&$     0.708$\\
 $(-16\div  -4)$&$ (  14\div  26)$&$        55.5  $&$     0.908$\\
 $(-26\div -14)$&$ (   9\div  21)$&$        65.1  $&$     1.247$\\
 $(-36\div -24)$&$ (   0\div  12)$&$        39.1  $&$     0.756$\\
 $(-10\div  10)$&$  (94\div 106)$&$        21.2  $&$     0.476$\\
 $(-10\div  10)$&$ (144\div 156)$&$        34.1  $&$     0.767$\\
\end{tabular}
\end{table}

\onecolumn

\end{document}